\documentclass[aps,10pt,tightenlines,amsmath,amssymb]{revtex4}
\usepackage{graphicx}
\usepackage{dcolumn}
\usepackage{bm}

\begin{document}

\title{Behavior of axion-like particles in smoothed out domain-like magnetic fields}

\author{Giorgio Galanti}
\email{gam.galanti@gmail.com}
\affiliation{INAF, Osservatorio Astronomico di Brera, Via Emilio Bianchi 46, I -- 23807 Merate, Italy}

\author{Marco Roncadelli}
\email{marco.roncadelli@pv.infn.it}
\affiliation{INFN, Sezione di Pavia, Via A. Bassi 6, I -- 27100 Pavia, Italy, and INAF}

\begin{abstract}
The existence of axion-like particles (ALPs) is predicted by many extensions of the standard model of elementary particles and in particular by theories of superstrings and superbranes. ALPs are very light, neutral, pseudo-scalar bosons which are supposed to interact with two photons. They can play an important role in high-energy astrophysics. Basically, in certain circumstances ALPs substantially enhance the photon survival probability $P_{\gamma \to \gamma} ({\cal E})$ of a beam emitted by a far-away source through the mechanism of photon-ALP oscillations (${\cal E}$ denotes the energy). But in order for this to work, an external magnetic field ${\bf B}$ must be present. In several cases ${\bf B}$ is modeled  as a domain-like network with `sharp edges': all domains have the same size $L_{\rm dom}$ (set by the ${\bf B}$ coherence length) and the same strength $B$, but the direction of ${\bf B}$ changes randomly and abruptly from one domain to the next. While this model has repeatedly been used so far since it greatly simplifies the calculations, it is obviously a highly mathematical idealization wherein the components of ${\bf B}$ are discontinuous across the edges (whence the name sharp edges). It is therefore highly desirable to go a step further, and to find out what happens when the edges are smoothed out, namely when the abrupt change of ${\bf B}$ is replaced by a smooth one. Moreover, this step becomes {\it compelling} when the photon-ALP oscillation length $l_{\rm osc}$ turns out to be comparable to -- or smaller than -- $L_{\rm dom}$, because in this case the photon survival probability $P_{\gamma \to \gamma} ({\cal E})$ critically depends on the domain shape. Finally, it would be more realistic to have $L_{\rm dom}$ randomly changing within a given range, since it looks rather unlikely that the coherence length of ${\bf B}$ should be everywhere the same especially when its source is sufficiently extended. In the present paper we propose a smoothed out version of the previous domain-like structure of ${\bf B}$ which incorporates the above changes, and we work out its implications. Even in the present case we are able to solve {\it analytically and exactly} the propagation equation of a monochromatic photon/ALP beam of energy ${\cal E}$ inside a single smoothed out  domain, thereby ultimately evaluating the corresponding photon survival probability $P_{\gamma \to \gamma} ({\cal E})$ exactly. This fact has the great advantage to drastically shorten the computation time in the applications involving computer simulations as compared to a numerical solution of the beam propagation equation. Actually, it turns out that the condition $l_{\rm osc} \lesssim L_{\rm dom}$ takes place when a photon/ALP beam of either very low ${\cal E}$ or very large ${\cal E}$ -- both in the gamma-ray band -- crosses a variety of astronomical objects, like radio lobes of flat spectrum radio quasars, spiral galaxies, starburst galaxies, elliptical galaxies and extragalactic space. Thus, the use of our model becomes {\it compelling} in all these instances, since the sharp edges model would yield unphysical results. The case of extragalactic space is of particular importance in view of the new generation of gamma-ray observatories like CTA, HAWC, GAMMA-400, LHAASO and TAIGA-HiSCORE, since in such a situation $l_{\rm osc} \lesssim L_{\rm dom}$ occurs for ${\cal E} \gtrsim {\cal O} (40 \, {\rm TeV})$  with a large uncertainty, depending on the choice of the model parameters (we shall come back to this fundamental issue in a subsequent paper).
\end{abstract}


\maketitle



\section{Introduction}

Axion-like particles (ALPs) are very light, neutral, pseudo-scalar bosons (for a review, see~\cite{alp1,alp2}). They are quite similar to the {\it axion}, namely the pseudo-Goldstone boson associated with the global Peccei-Quinn symmetry ${\rm U}(1)_{\rm PQ}$ proposed as a natural solution to the strong CP problem (for a review, see~\cite{axionrev1,axionrev2,axionrev3,axionrev4}). The axion interacts with fermions, two gluons and two photons, and one of its characteristic feature is the existence of a strict linear relationship between its mass $m$ and two-photon coupling constant $g_{a \gamma \gamma}$, which reads
\begin{equation}
\label{ma1}
m = 0.7 \, k \, \Bigl(g_{a \gamma \gamma} \,10^{10} \, {\rm GeV} \Bigr) \, {\rm eV}~,
\end{equation}
where $k$ is a model-dependent constant ${\cal O} (1)$~\cite{cgn1995}. However, ALPs differ from axions in two respects. In fact, their mass $m_a$ and two-photon coupling constant $g_{a \gamma \gamma}$ are {\it unrelated} parameters. Moreover, what matters for ALPs is their interaction with two photons. Additional interactions may be present but are not interesting for our purposes and will be discarded here, while for the axion its coupling to fermions and two gluons is essential in order for the Peccei-Quinn mechanism to work. 

\begin{figure}[h]
\centering
\includegraphics[width=0.40\textwidth]{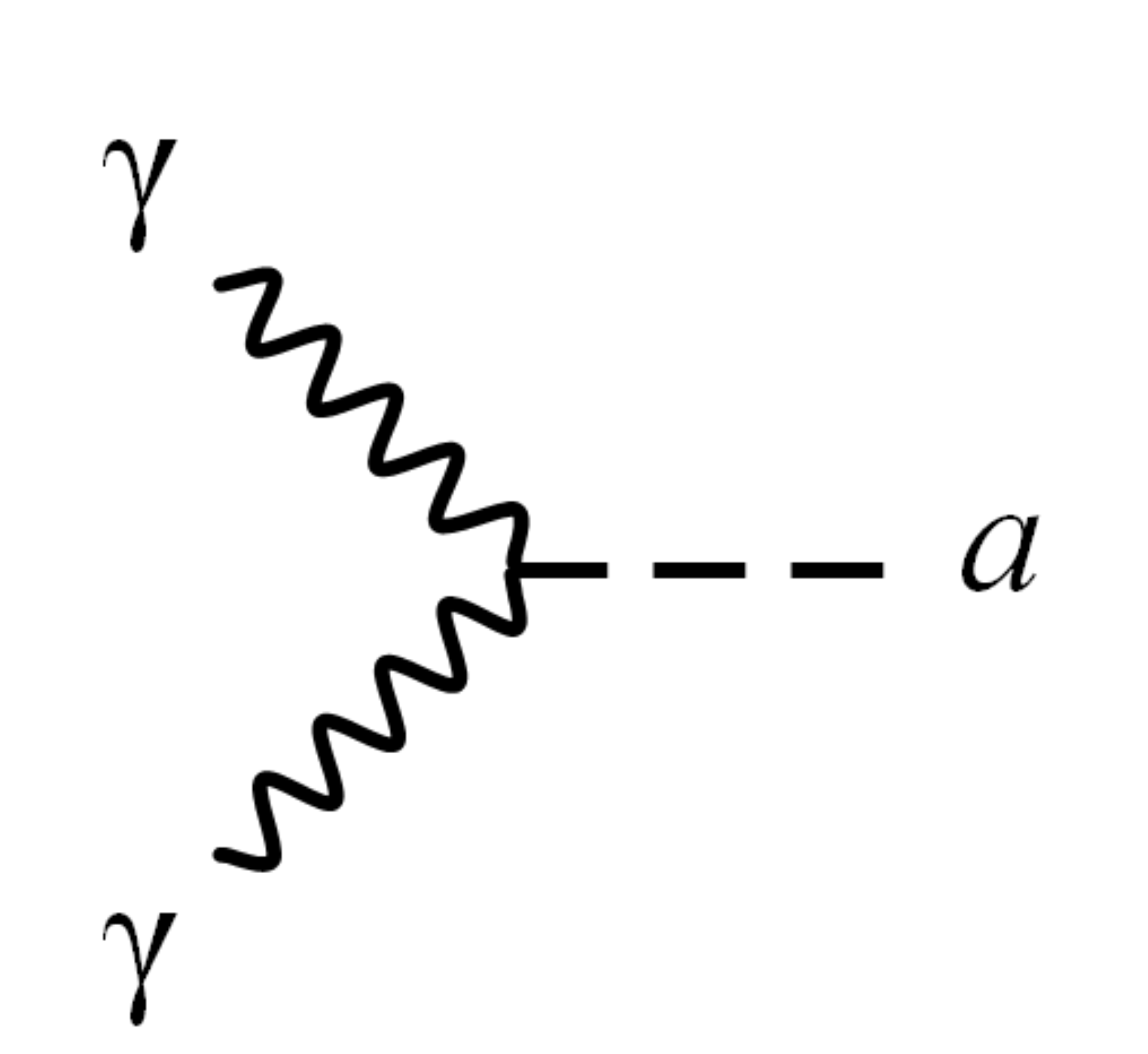}
\caption{\label{immagine3} Photon-photon-ALP vertex with coupling constant $g_{a \gamma \gamma}$.}
\end{figure}

So, we shall focus our attention on the two-photon-ALP coupling alone -- which is represented by the Feynman diagram in Figure~\ref{immagine3} -- denoting henceforth the ALP by $a$ for simplicity. 

\begin{figure}[h]
\centering
\includegraphics[width=.50\textwidth]{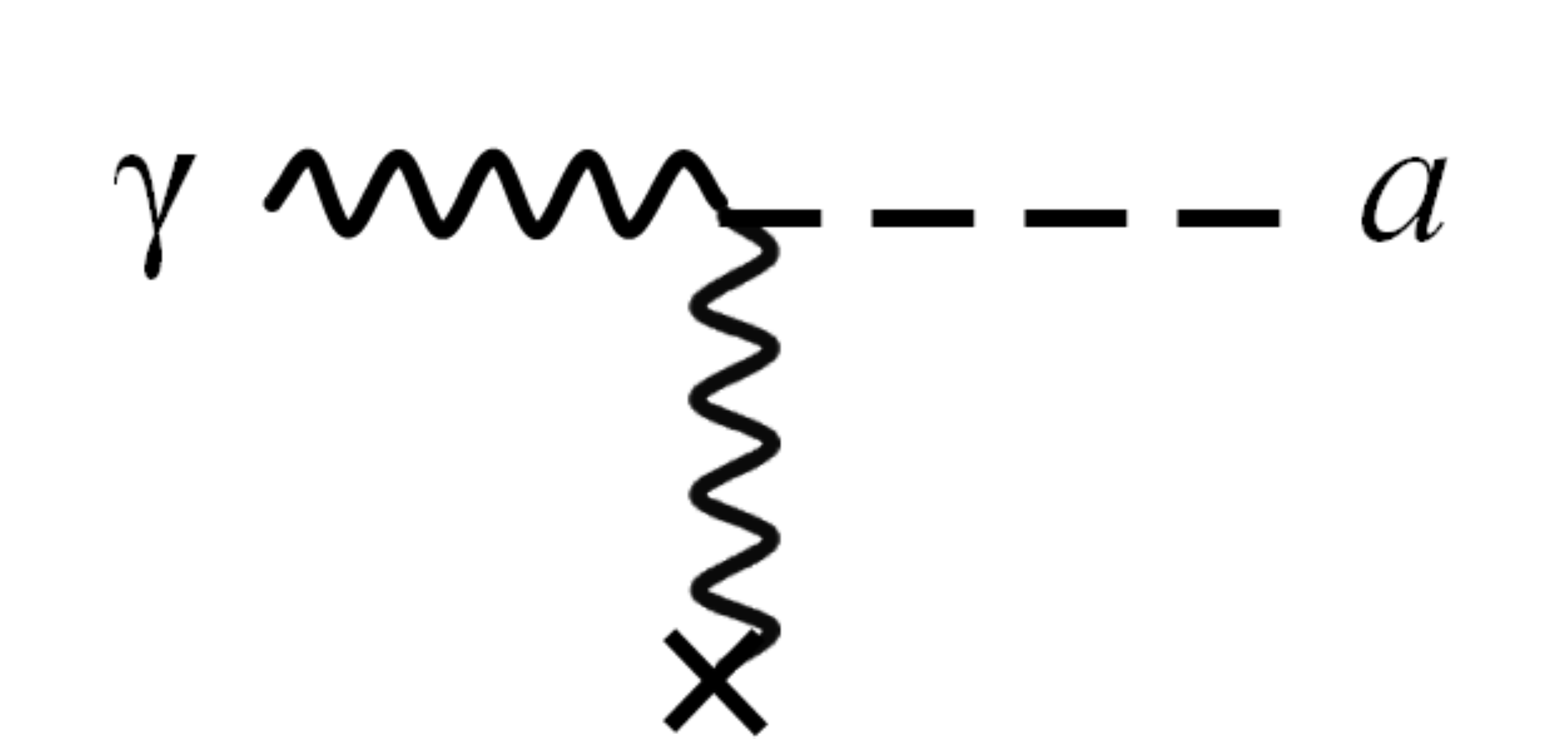}
\caption{\label{immagine4} $\gamma \to a$ conversion in the external magnetic field ${\bf B}$.}
\end{figure}

In the last few years, ALPs have attracted growing interest for a number of reasons. They are a generic prediction of many extensions of the standard model, notably of those based on the M theory, which encompasses superstring and superbrane theories (for an incomplete sample of references,  see ~\cite{turok1996,string1,string2,string3,string4,string5,axiverse,abk2010,cicoli2012,cicoli2013,dias2014,cicoli2014a,conlon2014,cicoli2014b,conlon2015,cicoli2017,scott2017,
concon2017,cisterna1,cisterna2}. In addition, depending on $m_a$ and $g_{a \gamma \gamma}$ they can be quite good candidates for cold dark matter~\cite{preskill,abbott,dine,arias2012}. Finally -- with  $m_a$ and $g_{a \gamma \gamma}$ in an appropriate range -- ALPs can give rise to very interesting astrophysical effects (for an incomplete sample of references, see~\cite{khlopov1992,khlopov1994,khlopov1996,raffelt1996,masso1996,csaki2002a,csaki2002b,cf2003,dupays,mr2005,fairbairn,mirizzi2007,drm2007,bischeri,sigl2007,dmr2008,shs2008,dmpr,mm2009,crpc2009,cg2009,bds2009,prada1,bronc2009,bmr2010,mrvn2011,prada2,gh2011,pch2011,dgr2011,frt2011,hornsmeyer2012,hmmr2012,wb2012,hmmmr2012,trgb2012,friedland2013,wp2013,cmarsh2013,mhr2013,hessbound,gr2013,hmpks,mmc2014,acmpw2014,rt2014,wb2014,hc2014,mc2014,bw2015,payez2015,trg2015,bgmm2016,fermi2016,day2016,sigl2016,sigl2017a,mcdmarsh2017,bnt2017,vlm2017,marshfabian2017,bnt2017a,zung2018,pani2018,zhang2018,mnt2018}), which are ultimately traced back to the fact that $\gamma \to a$ and $a \to \gamma$ conversions take place provided that an external magnetic (electric) field is present, as shown in Figure~\ref{immagine4}~\cite{sikivie,anselmo,rs1988}.

\begin{figure}[h]    
\centering
\includegraphics[width=0.40\textwidth]{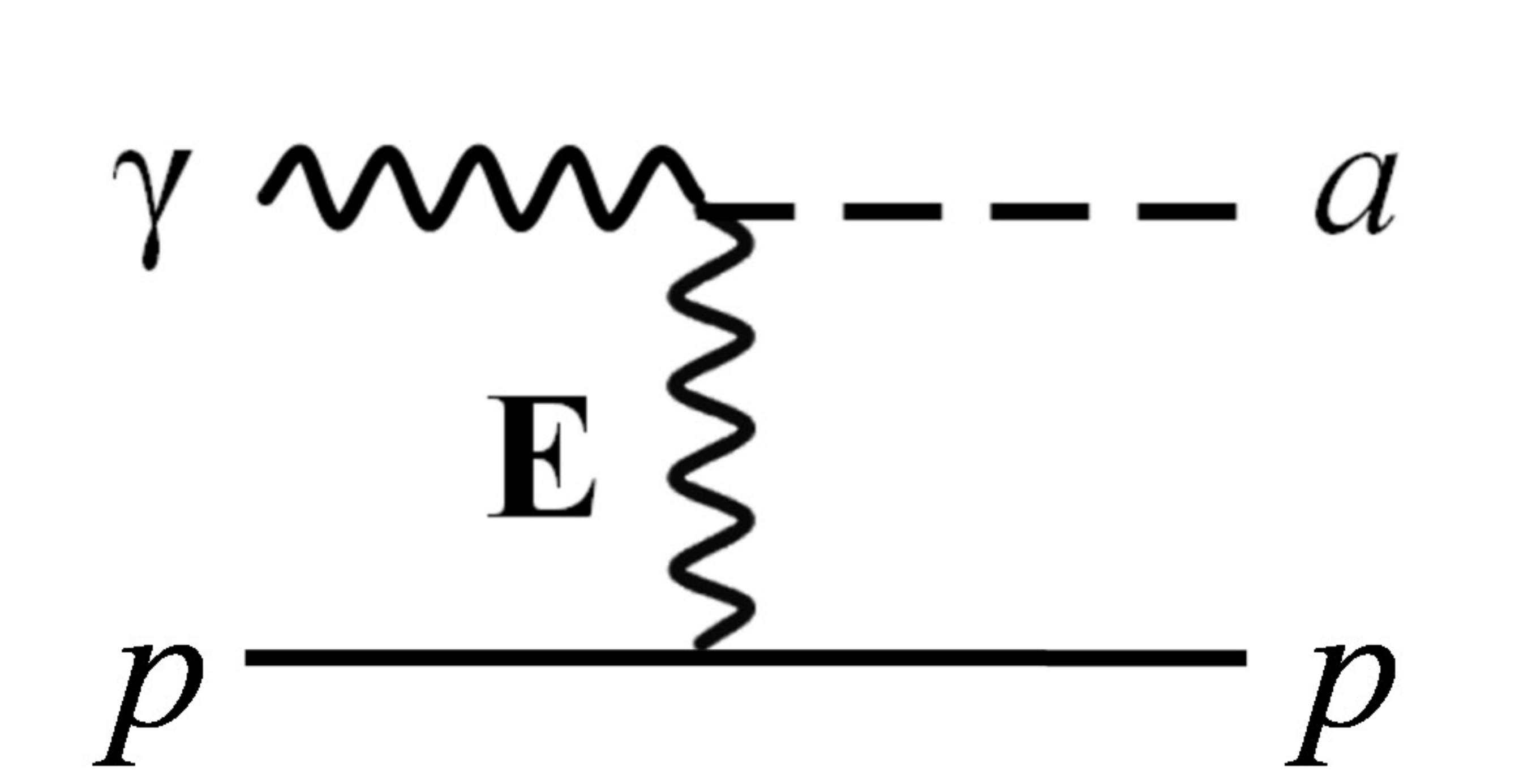}
\caption{Feynman diagram for the Primakoff process.}
\label{primakoff}
\end{figure}

Throughout this paper, we shall be concerned with the last point alone. Actually, the possibility of $\gamma \to a$ and $a \to \gamma$ conversions is of paramount importance for the ALP detection with {\it various} techniques. They include ALPs production in the Sun core through the Primakoff process $\gamma + {\rm ion} \to a + {\rm ion}$ -- represented in Figure~\ref{primakoff} -- and their subsequent conversion into photons of the same energy inside one or more superconducting hollow blind magnets pointing towards the Sun, like in the CAST (CERN Axion Solar Telescope) experiment~\cite{cast} and in the planned IAXO (International Axion Observatory) experiment~\cite{iaxo}. 

Remarkably, just the same conversion-reconversion mechanism allows for the `shining light through a wall' detection strategy like in the ALPS II experiment at DESY~\cite{alpsII} and in the planned STAX experiment~\cite{stax}. Furthermore, ALPs can be detected with other strategies developed by Avignone and collaborators~\cite{avignone1,avignone2,avignone3}. Finally, in case ALPs build up the bulk of the dark matter then they can also be discovered by the planned ABRACADABRA experiment~\cite{abracadabra}. Needless to say, in all these cases the magnetic field extends over a very limited distance, and in order to maximize the effect one has to resort to a magnetic field which is as strong as possible.    

Very interesting implications of the same conversion-reconversion phenomenon can occur in astrophysics, where the situation is usually just opposite: depending on the specific astronomical object crossed by a photon/ALP beam, the magnetic field can be very small as compared to the laboratory standards (apart from inside or around active galaxies, white dwarfs, protoneutron and neutron stars), but it extends over very much larger distances. Because the $\gamma \to a$ and $a \to \gamma$ conversion probability depends on the product of the magnetic field times the distance (as we shall see later), even a tiny magnetic field can produce sizable effects over astronomical distances, as stressed in~\cite{drm2007,dmr2008}.

\begin{figure}[h]
\centering
\includegraphics[width=1.0\textwidth]{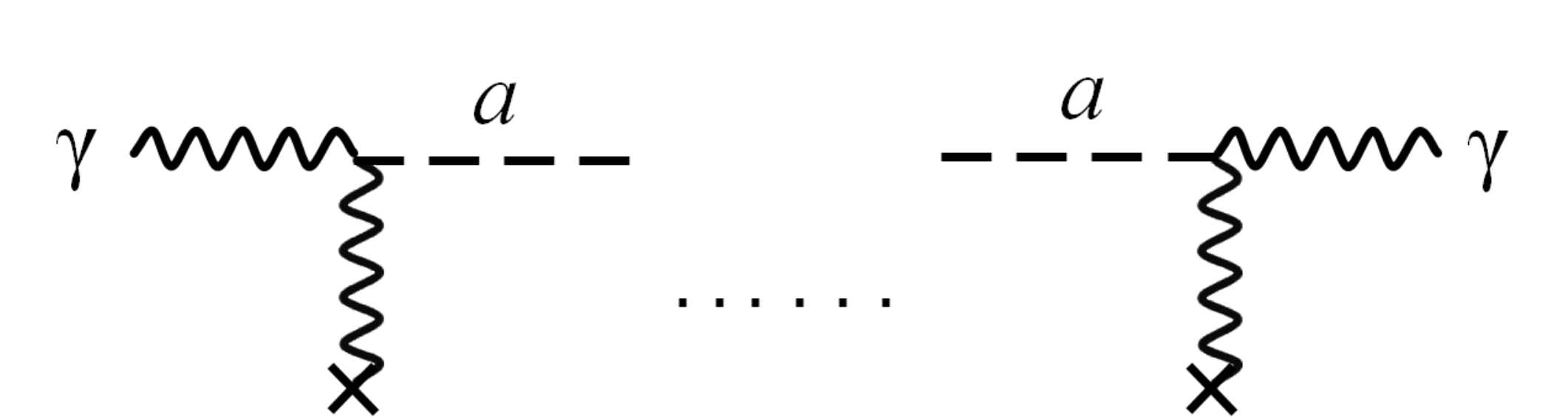}
\caption{\label{fey1} Schematic view of a $\gamma \leftrightarrow a$ oscillation in the external magnetic field ${\bf B}$.}
\end{figure}

Manifestly, as a photon propagates $\gamma \to a$ and $a \to \gamma$ conversions give rise to $\gamma \leftrightarrow a$ {\it oscillations} -- illustrated in Figure~\ref{fey1} -- which are quite similar to flavor oscillations of massive neutrinos apart from the need of the external magnetic field ${\bf B}$ in order to compensate for the spin mismatch~\cite{sikivie,anselmo,rs1988}. As a consequence, while flavor neutrino oscillations freely occur, the case of $\gamma \leftrightarrow a$ oscillations is more involved and will be discussed in Section III. Here, we merely state that the $\gamma \to a$  conversion probability is maximal and independent of the ALP mass $m_a$ and of its energy ${\cal E}$ for ${\cal E} > {\cal E}_L$ defined as
\begin{equation}
\label{eqprop1}
{\cal E}_L \equiv \frac{|m_a^2 - \omega^2_{\rm pl}|}{2 g_{a \gamma \gamma} \, B_T}~,
\end{equation}
where $\omega_{\rm pl}$ is the plasma frequency of the medium and $B_T$ will be defined below~\cite{bischeri,dmr2008,rem1}.

A natural question arises: how large is ${\cal E}_L$? Obviously the answer depends on $g_{a \gamma \gamma}$, but in order to get an orientation it is enough to know its upper bound. Owing to the negative result of the CAST experiment, the resulting bound is $g_{a \gamma \gamma} < 0.66 \cdot 10^{- 10} \, {\rm GeV}^{- 1}$ for $m < 0.02 \, {\rm eV}$ at the $2 \sigma$ level~\cite{cast}. Incidentally, exactly the same bound at the same confidence level has been obtained from the study of a particular kind of globular cluster stars~\cite{straniero}. Correspondingly -- neglecting $\omega_{\rm pl}$ for simplicity --  condition (\ref{eqprop1}) becomes~\cite{conventions}
\begin{equation}
\label{08012018a}
{\cal E}_L > 3.90 \cdot 10^{11} \left(\frac{m_a}{{\rm eV}} \right)^2 \left(\frac{{\rm G}}{B_T} \right) {\rm GeV}~.
\end{equation}
Thus, we see that for astrophysical magnetic fields in the range $\bigl(10^{- 10} - 10^{- 5} \bigr) \, {\rm G}$ (more about this, in Section II) even for very small values of $m_a$ it turns out that ${\cal E}_L$ tends to fall into the X- and $\gamma$-ray bands. Therefore, 
$\gamma \leftrightarrow a$ oscillations are expected to show up in {\it high-energy} astrophysics. As a consequence, denoting by ${\cal E}$ the photon energy, we will focus our attention throughout this paper on the regime ${\cal E} \gg m_a$.

\begin{figure}[h]
\centering
\includegraphics[width=.60\textwidth]{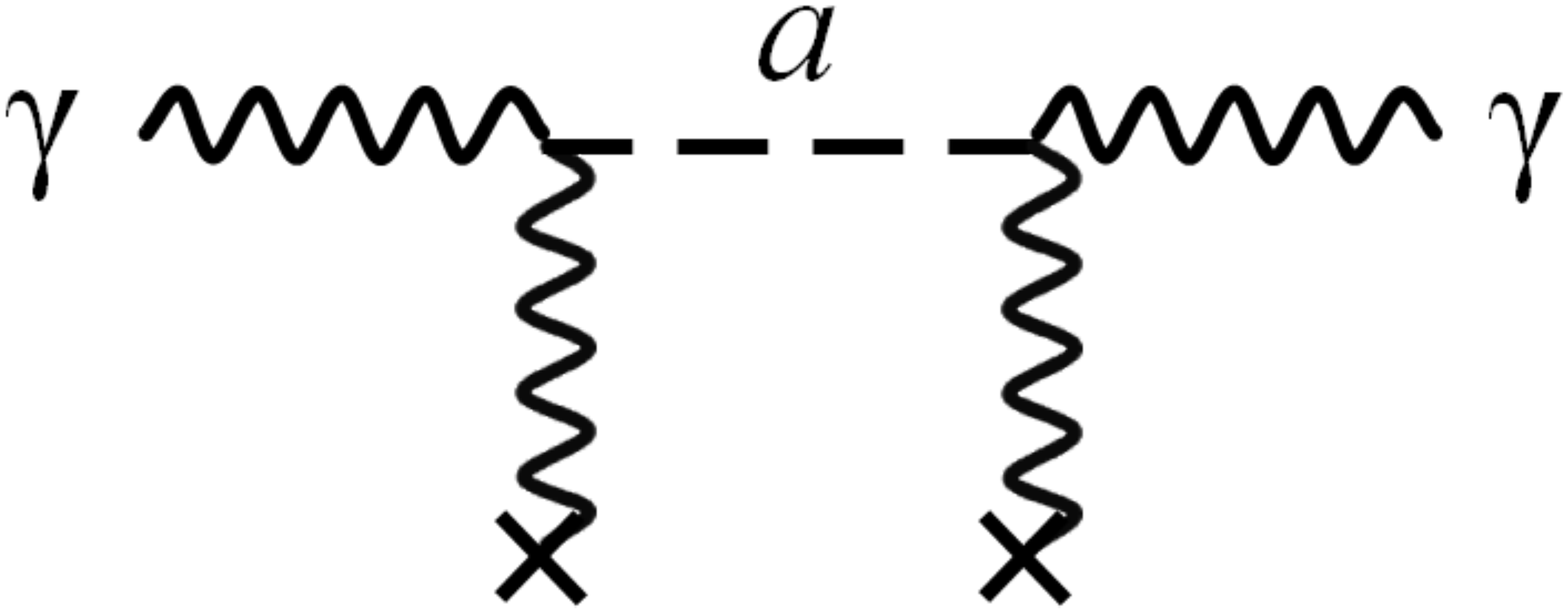}
\caption{\label{immagine5} Feynman diagram responsible for birefringence in the external magnetic field ${\bf B}$.}
\end{figure}

This is however not the end of the story. Because the $\gamma \gamma a$ vertex is $g_{a \gamma \gamma} \, a \, {\bf E} \cdot {\bf B}$, in the presence of an external magnetic field ${\bf B}$ only the component ${\bf B}_T$ in a generic plane $\Pi (y)$ orthogonal to the photon momentum ${\bf k}$ -- supposed along the $y$-axis -- couples to ALPs. Moreover, photons $\gamma_{\perp}$ with linear polarization orthogonal to the plane $\Pi_*$ defined by ${\bf k}$ and ${\bf B}$ do {\it not} mix with $a$, and so only photons $\gamma_{\parallel}$ with linear polarization inside $\Pi_*$ do mix with $a$. Hence, the term $g_{a \gamma \gamma} \, a \, {\bf E} \cdot {\bf B}$ act as a polarizer. Specifically -- besides $\gamma \leftrightarrow a$ oscillations -- for an initially linearly polarized photon beam two effects occur. One is {\it birefringence}, namely the linear polarization becomes elliptical with its major axis parallel to the initial polarization: this arises from the Feynman diagram in Figure~\ref{immagine5}. The other effect is {\it dichroism}, namely the polarization-dependent selective photon conversion, which causes the ellipse  major axis to become misaligned with respect to the initial polarization: this comes about from the Feynman diagram in Figure~\ref{immagine4}~\cite{mpz1986,rs1988,bmr2010,mrvn2011,pch2011,day2018}. 

So far, $\gamma \leftrightarrow a$ oscillations have been considered in a variety of astronomical objects, like active galaxies, spiral and elliptical galaxies, the Milky Way, clusters of galaxies and extragalactic space, since they are all magnetized structures. Apart from the case of active galaxies and clusters of galaxies, in all the other systems -- to be denoted by ${\cal T}$ for notational simplicity -- the magnetic field ${\bf B}$ can be modeled in first approximation according to the following procedure.

Throughout this paper we consider a high-energy monochromatic photon/ALP beam of energy ${\cal E}$. Suppose now that the beam crosses ${\cal T}$ along the $y$ direction, and we denote by $y_{\rm in} \equiv y_0$ and $y_{\rm ex} \equiv y_N$ the values of $y$ where the beam enters ${\cal T}$ and exits from ${\cal T}$, respectively. Whenever ${\bf B}$ has nearly the same strength in the region crossed by the beam and its coherence length is {\it very much smaller than the size} of ${\cal T}$ -- namely of $\bigl|y_{\rm ex} - y_{\rm in} \bigr|$ -- it has become customary to suppose that the ${\bf B}$ structure is a domain-like network: all {\bf $N$} magnetic domains have the same size $L_{\rm dom}$ equal to the ${\bf B}$ coherence length and in each domain ${\bf B}$ is assumed to be homogeneous, but that its direction changes {\it randomly} and {\it discontinuously} passing from one domain to the next (for a review, see~\cite{kronberg1994,grassorubinstein}). Because of the latter fact, this model will be referred to as having {\it sharp edges}. It is very important to emphasize that only a {\it single} realization of the beam propagation process is {\it observable}, since the deflection angles are random variables. Clearly, this picture captures the main feature of ${\bf B}$, namely its coherence over a single $n$-th domain {\it only}  ($1 \leq n \leq N$). Such a {\it domain-like sharp-edges} (DLSHE) model is of course a highly mathematical idealization, but since no discontinuity shows up in the photon survival probability $P^{\rm ALP}_{\gamma \to \gamma} \bigl({\cal E}; y_{\rm ex}, y_{\rm in} \bigr)$ across ${\cal T}$, a great simplification comes about since the beam propagation equation can easily be solved in a generic $n$-th domain where ${\bf B}$ is {\it homogeneous}. 

Yet, even in all those cases in which the DLSHE model turns out to be well motivated from a physical standpoint, it can nevertheless yield unphysical results. To see this, let us denote by $l_{\rm osc}$ the $\gamma \leftrightarrow a$ oscillation length and by $P_{\gamma \to a} (L_{\rm dom})$ the photon-ALP conversion probability across a single domain, and we argue as follows. Suppose first that $l_{\rm osc} \gg \, L_{\rm dom}$. Correspondingly, only a small fraction of a single oscillation is {\it coherently} affected by ${\bf B}$ in a single domain, and so $P_{\gamma \to a} (L_{\rm dom})$ becomes insensitive to the {\it shape} of the magnetic domain: in this case the DLSHE model can be successfully applied.

On the other hand, when $l_{\rm osc} \lesssim L_{\rm dom}$ a whole oscillation -- or even several oscillations -- are contained inside a single domain, thereby implying that 
$P_{\gamma \to a} (L_{\rm dom})$ now {\it depends} on the domain {\it shape}. This is indeed what happens above or below a certain energy threshold in a variety of astronomical objects -- radio lobes, spiral galaxies, starburst galaxies, elliptical galaxies {\it and}  extragalactic space -- as we will see later. For instance, a particular case study in which indeed the condition $l_{\rm osc} \lesssim L_{\rm dom}$ occurs -- owing to the photon dispersion on the CMB (Cosmic Microwave Background)~\cite{raffelt2015} -- is a photon/ALP beam of energy ${\cal E} \gtrsim {\cal O} (40 \, {\rm TeV})$ with a large uncertainty traveling in extragalactic space~\cite{raffeltvogel,paper2}. Manifestly, whenever $l_{\rm osc} \lesssim L_{\rm dom}$ the use of the DLSHE model gives rise to meaningless results because of its unphysical nature. In addition, whenever ${\cal T}$ is a rather extended astronomical object, il looks fairly unrealistic to suppose that all magnetic domains have exactly the same size: a random spread of $L_{\rm dom}$ within a given range looks much more realistic.

In order to save the situation -- that is to say, to retain the advantages of the DLSHE model while still getting physically meaningful results -- we have to smooth out the edges in such a way that the components of ${\bf B}$ change {\it continuously} across the edges between adjacent domains, thereby obtaining the {\it domain-like smooth-edges} (DLSME) model. Moreover, we allow for the possibility of a random change of the domain size within a fixed range. Depending on the amount of smoothing, the DLSME model can indeed be regarded as a realistic description. Clearly, the generalization of the DLSHE model to the DLSME model brings about the difficulty of solving the beam propagation equation inside a single smoothed out domain where ${\bf B}$ is {\it not anymore homogeneous}. Nevertheless, by a somewhat clever use of the Laplace transform we are able to solve such an equation 
{\it analytically and exactly} for an arbitrary amount of smoothing. 

\

The paper is organized as follows. Section II discusses the application of domain-like magnetic field models of the considered kind to various astronomical objects, including extragalactic space. Section III briefly recalls the formalism whereby the photon survival probability $P^{\rm ALP}_{\gamma \to \gamma} \bigl({\cal E}; y_{\rm ex}, y_{\rm in} \bigr)$ is computed within domain-like magnetic field models in the presence of $\gamma \leftrightarrow a$ oscillations. Section IV presents a discussion of the three main regimes defined by the ALP energy, and for each of them we report the approximated values of the $\gamma \leftrightarrow a$ oscillation length and of the $\gamma \to a$ conversion probability in a single domain. Section V describes our smoothing proposal, which basically consists in taking the orientation angles $\phi (y)$ and $\theta (y)$ of ${\bf B} (y)$ in each domain constant in the inner part and linearly increasing or decreasing across the edges in such a way that they smoothly connect with the values in the two adjacent domains. Moreover, we show that the three-dimensional problem can be reduced to a two-dimensional one inside a single domain, and we also explain how to evaluate the photon survival probability along a single realization of a photon/ALP beam propagating in the $y$-direction emitted by a compact source and reaching us, since -- as already emphasized -- it is indeed a single realization that is actually observed. In Section VI we solve the beam propagation equation in the domain region where $\phi = {\rm constant}$, while in Section VII we address the more difficult task of solving exactly and analytically the beam propagation equation in the region where $\phi \neq {\rm constant}$. In Section VIII we compare the cases $l_{\rm osc} > L_{\rm dom}$ and $l_{\rm osc} < L_{\rm dom}$ and investigate the effect of smoothing of the photon survival probability $P^{\rm ALP}_{\gamma \to \gamma} \bigl({\cal E}; y_{\rm ex}, y_{\rm in} \bigr)$. In Section IX we determine the energy at which the DLSHE model must be replaced by our DLSME model for some of the astronomical objects considered in Section II. Finally, we offer our conclusions in Section X.

\section{Varieties of astrophysical magnetic field}

As we said, we shall be concerned with a high-energy monochromatic photon/ALP beam of energy ${\cal E}$ emitted by a compact astronomical object like a gamma ray burst or a blazar. We recall that nearly one-tenth of Active Galactic Nuclei (AGNs) have an accretion disk and two jets emanating from the centre and perpendicular to the disk, in which photons from the central engine are accelerated to very-high-energy (VHE, $100 \, {\rm GeV} \lesssim {\cal E} \lesssim 100 \, {\rm TeV}$). With the presently available capabilities, the photon beam emitted by one jet can be detected only when it happens to be occasionally oriented along the line of sight: in this  case the AGN is called a {\it blazar}. There exist two different kinds of blazars: BL Lacs objects and Flat Spectrum Radio Quasars (FSRQs). A feature of FSRQs -- which is not shared by BL Lacs -- is the existence of a magnetized radio lobe with hot spots at the end of a jet. As a rule, FSRQs are considerably stronger VHE photon sources as compared to BL Lacs~\cite{conventions}. 

So, the considered beam necessarily crosses the host galaxy, extragalactic space and the Milky Way. Whenever it is emitted by the jet of a FSRQ it also traverses a radio lobe. Moreover, it can also cross a spiral or elliptical galaxy or a cluster of galaxies if they turn 
out to be on the line of sight. Because they are all magnetized structures, they can give 
rise to $\gamma \leftrightarrow a$ oscillations in the beam. 

As stressed in the previous Section, in first approximation the magnetic field ${\bf B}$ generated by an astronomical objects ${\cal T}$ can be described by a domain-like model provided that the above assumptions are met. Below, we consider some specific cases which are of interest for us. Note that we do not consider the blazar jets because in this case a domain-like ${\bf B}$ model does not work as well as galaxy clusters where 
$B$ is different in different domains.

\

\noindent {\it Radio lobes and hot spots} -- It is impossible to estimate the magnetic fields in these structures in a model-independent fashion. According to the model of Ghisellini {\it et al}.~\cite{ghisello2014a}, radio lobes have a spherical shape of radius $R_{\rm lobe} \simeq 50 \, {\rm kpc}$ homogeneously filled by a magnetic field ${\bf B}_{\rm RL}$ which varies considerably from source to source and lies within the range $2.3 \, \mu{\rm G} \lesssim B_{\rm RL} \lesssim 13 \, \mu{\rm G}$ with coherence length of about $10 \, {\rm kpc}$. In the hot spot, the magnetic field $B_{\rm HS}$ is still larger, in the range $16.5 \, \mu{\rm G} \lesssim B_{\rm HS} \lesssim 169 \, \mu{\rm G}$ but the coherence length is by far too small to be an interesting quantity. Hence, in first approximation radio lobes can be described by a domain-like model with the above values of $B$ and the coherence length.

\

\noindent {\it Milky Way} -- Nowadays the magnetic field of the Galaxy is fairly well known, and the two most detailed models have been developed by Pshirkov, Tinyakov, Kronberg and Newton-McGee~\cite{ptkn2011} and by Jansson and Farrar~\cite{jf2012}. Basically, the main difference between them is that in the first model the magnetic field is more concentrated towards the disk. Nevertheless, when simple estimates are sufficient, the Galactic magnetic field can be represented by two components which both lie within the disk: a regular one described by a domain-like model with $L_{\rm dom} \simeq 10 \, {\rm kpc}$ and ${\rm B} \simeq 5 \, \mu{\rm G}$ and a turbulent one having a Kolmogorov spectrum with index $\alpha = - 5/3$~\cite{hfn2004}. 

\

\noindent {\it Spiral galaxies} --  Because the Milky Way is a typical bright spiral galaxy, it is natural to suppose that a somewhat similar magnetic field is present also in other spiral galaxies. And to the extent that observations are possible, they confirm such an expectation~\cite{vallee2011}. For instance, in our neighboring spirals M31 and M33 it is found for the regular component a domain-like structure $L_{\rm dom} \simeq 10 \, {\rm kpc}$ and ${\rm B} \simeq 6 \, \mu{\rm G}$.  As a consequence, also in general spirals  the large-scale magnetic field can be described in first approximation by a domain-like model with $L_{\rm dom} \simeq 10 \, {\rm kpc}$. As far as the ordered $B$ is concerned, it depends on the particular kind of spiral. In massive spirals we have ${\rm B} \simeq 10 \, \mu{\rm G}$ on average, which increases up to $B = (10 - 15) \, \mu{\rm G}$ in gas rich spirals with high star-formation rate~\cite{beck2016}, and up to ${\rm B} \simeq 50 \, \mu{\rm G}$ in starburst galaxies~\cite{kronbergbook}.

\

\noindent {\it Elliptical galaxies} -- Unfortunately, very little is known about the magnetic field in this kind of galaxies. Nevertheless, it has been argued that supernova explosions and stellar motion give rise to a turbulent ${\bf B}$ which can be described by a domain-like model with average strength ${\rm B}  \simeq 5 \, \mu{\rm G}$ and $L_{\rm dom} \simeq 150 \, {\rm pc}$~\cite{moss1996}. 

\

\noindent {\it Extragalactic space} -- Unfortunately, the morphology and strength of the extragalactic magnetic field ${\bf B}_{\rm ext}$ is totally unknown, and so it is not surprising that various very different configurations of ${\bf B}_{\rm ext}$ have been proposed~\cite{kronberg1994,grassorubinstein,wanglai,jap}. It goes without saying that the final result -- namely the photon survival probability $P^{\rm ALP}_{\gamma \to \gamma} \bigl({\cal E}; y_{\rm ex}, y_{\rm in} \bigr)$ for a beam emitted by a far away compact astronomical object and reaching us -- changes greatly depending on the ${\bf B}_{\rm ext}$ morphology. Yet, the strength of ${\bf B}_{\rm ext}$ is constrained to lies in the range $10^{- 7} \, {\rm nG} \lesssim {\rm B}_{\rm ext} \lesssim 1.7 \, {\rm nG}$ on the scale  ${\cal O} (1 \, {\rm Mpc})$~\cite{neronov2010,durrerneronov,pshirkov2016,sigl2017}.

Nevertheless, also in this context the above considered domain-like structure of 
${\bf B}_{\rm ext}$ has commonly been used so far. Still -- at variance with the previous situations -- this kind of morphology is not simply a first approximation but it relies upon a realistic physical model involving energetic {\it galactic outflows}. Accordingly, ionized galactic matter is ejected into extragalactic space with the magnetic field frozen in, which gets amplified by turbulence and magnetizes the surrounding space. This idea was first put forward in 1968 by Rees and Setti~\cite{reessetti} and in 1969 by Hoyle~\cite{hoyle} in their attempts to modeling radio sources. A different and very interesting situation has been contemplated in 1999 by Kronberg, Lesch and Hopp~\cite{kronberg1999}. They proposed that dwarf galaxies are ultimately the source of ${\bf B}_{\rm ext}$. Specifically, they start from the clear consensus that supernova-driven galactic winds are a crucial ingredient in the evolution of dwarf galaxies. Next, they show that shortly after a starburst, the kinetic energy supplied by supernovae and stellar winds inflate an expanding superbubble into the surrounding interstellar medium of a dwarf galaxy. Moreover, they demonstrate that the ejected thermal and cosmic-ray gas -- significantly magnetized -- will become mixed into the surrounding intergalactic matter, which will coexpand with the universe. This picture leads to ${\rm B}_{\rm ext} = {\cal O} ( 1 \, {\rm nG})$ on the scale ${\cal O} (1 \, {\rm Mpc})$. Actually, in order to appreciate the relevance of dwarf galaxies, it is useful to recall that our Local Group -- which is dominated by the Milky Way and Andromeda -- contains 38 galaxies, 23 of which are dwarfs. Because the Local Group has nothing special, it follows that dwarf galaxies are roughly 10 times more abundant than bright Hubble type galaxies. The considered result is in agreement with observations of Lyman-alpha forest clouds~\cite{cowie1995}. A similar situation was further investigated in 2001 by Furlanetto and Loeb~\cite{furlanettoloeb} in connection with quasars outflows. What about ${\bf B}_{\rm ext}$ and $L_{\rm dom}$ in the present case? Also the scenario of Furlanetto and Loeb predicts ${\rm B}_{\rm ext} = {\cal O} ( 1 \, {\rm nG})$ on the scale ${\cal O} (1 \, {\rm Mpc})$. Moreover, also normal galaxies possess this kind of ionized matter outflows -- especially ellipticals and lenticulars -- due to the central AGN (see e.g.~\cite{ciotti} and references therein) and supernova explosions. Remarkably, this picture is in agreement with numerical simulations~\cite{bertone2006}. Uncontroversial evidence of galactic outflows comes from the high metallicity (including strong iron lines) of the intracluster medium of regular galaxy clusters, which are so massive that matter cannot escape. Manifestly, in all these models the seeds of ${\bf B}_{\rm ext}$ are galaxies, a fact which indeed explains the three main features of the considered domain-like model for ${\bf B}_{\rm ext}$: its cell-like morphology arising from their galactic origin, why ${\bf B}_{\rm ext}$ has nearly the same strength in all domains -- which means  around each galaxy -- and why the ${\bf B}_{\rm ext}$ direction changes randomly from the neighborhood of a galaxy to that of another one, since they are uncorrelated. Needless to say, the change of the ${\bf B}_{\rm ext}$ direction is smooth across the domain edges, and so once again the DLSME model is realistic whereas the DLSHE model is not. Unfortunately, it is still impossible to determine the strength of ${\bf B}_{\rm ext}$ in every domain, but the overall picture seems to suggest that it is nearly the same in all domains.

A radically different approach to the extragalactic magnetic field relies upon the magnetohydrodynamic cosmological simulations (see e.g.~\cite{vazza2014,vazza2016} 
and references therein). Basically, an initial condition for a cosmological ${\bf B}_{\rm ext}$ is chosen 
{\it arbitrarily} during the dark age and its evolution as driven by structure formation is investigated. The link with the real world is the request to reproduce cluster magnetic fields, which fixes {\it a posteriori} the initial condition of ${\bf B}_{\rm ext}$. As a by-product a prediction of the magnetic field ${\bf B}_{\rm fil}$ inside filaments in the present Universe emerges. However, this cannot be the whole story. Apart from leaving totally unanswered the question concerning the seed of primordial magnetic fields, what is missing are just galactic outflows. Specifically, this issue has a two-fold relevance: inside galaxy clusters and in extragalactic space. 

\begin{itemize}

\item As stressed above, galactic outflows are a reality in regular galaxy clusters. Actually, it has been claimed in 2009 by Xu, Li, Collins and Norman that the magnetic field ejected by a central AGN during the cluster formation can be amplified by turbulence during the cluster evolution in such a way to explain the observed cluster magnetic fields~\cite{xu2009}. Moreover, still in 2009 Donnert, Dolag, Lesch and M\"uller have found that the strength and structure of the magnetic fields observed in clusters of galaxies are well reproduced for a wide range of the model parameters by galactic outflows~\cite{donnert2009}. As a consequence, the requirement to reproduce cluster magnetic fields can totally mislead the expectations based on cosmological hydrodynamic simulations and in particular the present value of the magnetic field ${\bf B}_{\rm fil}$ inside filaments~\cite{vazza2014,vazza2016}.

\item Cosmological hydrodynamic simulations fail to simulate the effects of galactic outflows in extragalactic space simply because they are not included, in spite of the fact that their existence has never been in doubt. This fact adds a further uncertainty in the estimate of 
${\bf B}_{\rm fil}$.

\end{itemize}

Thus, we stick to the DLSME model throughout this paper.

\section{Setting the stage}

For our purposes, ALPs are described by the following Lagrangian
\begin{equation}
\label{lagr}
{\cal L}_{\rm ALP} =  \frac{1}{2} \, \partial^{\mu} a \, \partial_{\mu} a - \frac{1}{2} \, m_a^2 \, a^2 - \, \frac{1}{4 } g_{a\gamma\gamma} \, F_{\mu\nu} \tilde{F}^{\mu\nu} a = \frac{1}{2} \, \partial^{\mu} a \, \partial_{\mu} a - \frac{1}{2} \, m_a^2 \, a^2 + g_{a\gamma\gamma} \, {\bf E} \cdot {\bf B}~a~,
\end{equation}     
where ${\bf E}$ and ${\bf B}$ are the electric and magnetic components of the electromagnetic tensor $F_{\mu\nu}$ ($\tilde{F}^{\mu\nu}$ is its dual). 

We suppose hereafter that the monochromatic photon/ALP beam of energy ${\cal E}$ travels along the $y$ direction in a magnetic field, and so in Eq. (\ref{lagr}) ${\bf E}$ is the electric field of a beam photon while ${\bf B}$ is the external field generated by ${\cal T}$.  

Within this Section we confine our attention to a single magnetic domain, dropping the subindex $n$ for simplicity.  

We recall that we are in the regime ${\cal E} \gg m_a$, and so -- as first pointed out by Raffelt and Stodolsky~\cite{rs1988} -- the photon/ALP beam propagation equation becomes a Schr\"odinger-like equation with $t$ replaced by the coordinate $y$ along the beam. Whence 
\begin{equation}
\label{eqprop}
\left( i \frac{d}{d y} + {\cal E} + {\cal M} ({\cal E}, y) \right) \, \psi (y) = 0~,
\end{equation}
with
\begin{equation}
\label{smvett}
\psi ( y ) \equiv \left(
\begin{array}{c}
\gamma_1 ( y ) \\
\gamma_2 ( y ) \\
a( y ) \\
\end{array}
\right)~,
\end{equation}
where $\gamma_1 ( y )$ and $\gamma_2 ( y )$ are the photon amplitudes with polarization along the $x$- and $z$-axis, respectively, while $a ( y )$ is the ALP amplitude. This achievement is of great importance, since it allows the beam to be treated by means of the formalism of non-relativistic quantum mechanics.

Specifically, we decompose ${\bf B} (y)$ into a longitudinal component along the $y$-axis ${\bf B}_L (y)$ and a transverse component ${\bf B}_T (y)$ in the planes $\Pi (y)$ perpendicular to the $y$-axis. Then the mixing matrix ${\cal M} ({\cal E}, y)$ entering Eq. (\ref{eqprop}) has the form
\begin{equation}
\label{sm3}
{\cal M}({\cal E}, y ) \equiv
\left(
\begin{array}{ccc}
w ({\cal E},  y ) & 0 & v (y) \, s( y ) \\
0 & w ({\cal E},  y ) & v (y) \, c( y ) \\
v (y) \, s( y ) & v (y) \, c( y ) & u ({\cal E})
\end{array}
\right)~.
\end{equation} 
where $s( y ) \equiv {\rm sin} \, \phi( y )$, $c( y ) \equiv{\rm cos} \, \phi( y )$ with $\phi ( y )$ denoting the angle between ${\bf B}_T (y)$ and the fiducial fixed $x$-direction in the $\Pi  ( y )$ planes which is the {\it same} for all domains. We further denote by $\theta ( y )$ the angle between ${\bf B}$ and the $y$-axis, so that we have ${\bf B}_T ( y ) = {\bf B} (y) \, {\rm sin} \, \theta ( y )$. We stress that the longitudinal component ${\bf B}_L ( y ) = {\bf B} (y) \, {\rm cos} \, \theta ( y )$ does not couple to ALPs.

The quantities entering ${\cal M}({\cal E}, y )$, namely $v ( y )$ and $u ({\cal E})$ are real but $w ({\cal E}, y )$ can be complex. As a consequence, in general ${\cal M}({\cal E}, y )$ is not self-adjoint, thereby implying that the evolution operator is not unitary. So, the considered beam can formally be regarded as a {\bf {\it three-level non-relativistic unstable quantum system}}. We have written the matrix ${\cal M} ({\cal E}, y )$ in an abstract form for notational simplicity in view of our subsequent calculations, but the meaning of the various quantities is as follows 
\begin{equation}
\label{mr15112017b}
u ({\cal E}) \equiv - \, \frac{m_a^2}{2 {\cal E}}~, 
\end{equation}
\begin{equation}
\label{19042018a}
w_1 ({\cal E}) \equiv - \, \frac{\omega^2_{\rm pl}}{2 {\cal E}}
\end{equation}
\begin{equation}
\label{19042018b}
w_2 ({\cal E}, y) \equiv \left[1.42 \cdot 10^{- 4} \, \left(\frac{B_T ( y )}{B_{\rm cr}} \right)^2 + 0.522 \cdot 10^{- 42} \right] {\cal E}~,
\end{equation} 
\begin{equation}
\label{mr15112017a}
w ({\cal E}, y ) \equiv w_1 ({\cal E}) + w_2 ({\cal E}, y)  + \frac{i}{2 \lambda_{\gamma} ({\cal E})}~, 
\end{equation}
\begin{equation}
\label{mr15112017c}
v ( y ) \equiv \frac{g_{a \gamma \gamma} \, B_T ( y )}{2}~. 
\end{equation}
The term in $w_1 ({\cal E})$ accounts for the plasma frequency, the one in $w_2 ({\cal E},  y )$ takes into account the one-loop QED vacuum polarization ($B_{\rm cr} \simeq 4.41 \cdot 10^{13} \, {\rm G}$ is the critical magnetic field)~\cite{rs1988,he1936,we1936,schwinger,adler,notea} (first term) and the photon dispersion on the CMB~\cite{raffelt2015} (second term), while the imaginary term in $w ({\cal E},  y )$ is relevant when ${\cal T}$ absorbs photons: 
$\lambda_{\gamma} ({\cal E})$ is the corresponding mean free path.  

\

We proceed to apply Eq. (\ref{eqprop}) with ${\cal M} ({\cal E}, y)$ of the form (\ref{sm3}) 
to a generic domain-like model and inside a single domain. Recalling that the transfer matrix $U \bigl({\cal E}; y, y_0; \phi (y), \theta (y) \bigr)$ is the solution of Eq. (\ref{eqprop}) with initial condition $U \bigl({\cal E}; y_0, y_0; \phi (y_0), \theta (y_0) \bigr) = 1$, it turns out that the propagation of any wave function across one domain can be represented generically as
\begin{equation}
\label{trmatr}
\psi( y ) = U \bigl({\cal E}; y, y_0; \phi (y), \theta (y) \bigl)\, \psi(y_0)~.
\end{equation}
Moreover, by setting
\begin{equation}
\label{expon}
U \bigl({\cal E}; y, y_0; \phi (y), \theta (y) \bigr) = e^{ i {\cal E} \left( y - y_0 \right)} \, {\cal U} \bigl({\cal E}; y, y_0; \phi (y), \theta (y) \bigr)
\end{equation}
we find that ${\cal U} \bigl({\cal E}; y, y_0; \phi (y), \theta (y) \bigr)$ is the transfer matrix associated with the reduced Sch\"odinger-like equation
\begin{equation}
\label{redeqprop}
\left(i \frac{d}{d y} + {\cal M} ({\cal E}, y) \right) \psi( y ) = 0~.
\end{equation}

Within the present context, the wave function $\psi (y)$ describes a linearly polarized beam, but we have seen that a photon/ALP beam changes its polarization as it propagates, and so 
in order to describe an unpolarized beam it becomes compelling to employ the density matrix $\rho ( y )$, which obeys the Von Neumann-like equation associated with Eq. (\ref{redeqprop}), namely
\begin{equation}
\label{vneum}
i \frac{d \rho (y)}{d y} = \rho (y) \, {\cal M}^{\dag} ({\cal E}, y) - {\cal M} ({\cal E}, y) \, \rho (y)~,
\end{equation}
whose solutions can be represented in terms of ${\cal U} \bigl({\cal E}; y, y_0; \phi (y), \theta (y) \bigr)$ as
\begin{equation}
\label{unptrmatr}
\rho ( y ) = {\cal U} \bigl({\cal E}; y, y_0;\phi (y), \theta (y) \bigr) \, \rho_0 \, {\cal U}^{\dag} \bigl({\cal E}; y, y_0; \phi (y), \theta (y) \bigr)~.
\end{equation}
Hence, the probability that the beam in the initial state $\rho_0$ at $y_0$ will be found in the final state $\rho$ at $y$ reads
\begin{equation}
\label{unpprob}
P^{\rm ALP}_{\gamma \to \gamma} \bigl({\cal E}; y, \rho ; y_0,  \rho_0; \phi (y), \theta (y) \bigr) = {\rm Tr} \Bigl[\rho \, {\cal U} \bigl({\cal E}; y, y_0; \phi (y), \theta (y) \bigr) \, \rho_0 \, {\cal U}^{\dag} \bigl({\cal E}; y, y_0; \phi (y), \theta (y) \bigr) \Bigr]
\end{equation}
with ${\rm Tr} \, \rho_0 = {\rm Tr} \, \rho =1$~\cite{dgr2011}.

\section{Oscillation length and conversion probability}

Before proceeding further, it is very useful to discuss the oscillation length $l_{\rm osc} 
({\cal E}, y)$ and the conversion probability $P_{\gamma \to a} ({\cal E}, y)$ within a single domain. 

Two points should be stressed. First -- because absorption is independent of the $\gamma \leftrightarrow a$ oscillation length $l_{\rm osc} ({\cal E})$ -- it can be discarded in this Section. Second, in principle we should define $l_{\rm osc} ({\cal E})$ and $P_{\gamma \to a} ({\cal E}, y)$ by means of the transfer matrix ${\cal U} (E; y, y_0; \phi (y), \theta (y))$  to be derived later on. However, the resulting expressions would be so cumbersome to become effectively obsolete. As a consequence, we prefer to work within the DLSHE model, which means that we have to set $s (y) = 0$ and $c(y) = 1$ into Eq. (\ref{sm3}). Otherwise stated, we are following a perturbative approach. Correspondingly, we have~\cite{rs1988}
\begin{equation}
\label{07052017b}     
l_{\rm osc} ({\cal E}) \equiv \frac{2 \pi}{\left[\bigl(w ({\cal E}) - u ({\cal E}) \bigr)^2 + 4 \, v^2 \right]^{1/2}}~, 
\end{equation}
from which it can be shown that the $\gamma \to a$ conversion probability is
\begin{equation}
\label{07052017c}
P_{\gamma \to a} ({\cal E}, y) = \left(\frac{v \, l_{\rm osc} ({\cal E})}{\pi} \right)^2 {\rm sin}^2 \left(\frac{\pi y}{l_{\rm osc} ({\cal E})} \right)~, \ \ \ \ \ \ \ \ \ \ \ \ y \leq L_{\rm dom}~. 
\end{equation}

Now, for the reader's convenience we rewrite the oscillation length (\ref{07052017b}) and the conversion probability (\ref{07052017c}) in an explicit form, recalling the definitions (\ref{mr15112017b}), (\ref{19042018a}), (\ref{mr15112017a}) and (\ref{mr15112017c}). Hence, the {\it oscillation length} $l_{\rm osc} ({\cal E})$ becomes
\begin{equation}
\label{a17}
l_{\rm osc} ({\cal E}) = 2 \pi \left(\left\{\frac{m_a^2  - \omega_{\rm pl}^2 }{2 {\cal E}} + \left[1.42 \cdot 
10^{- 4} \, \left(\frac{B_T}{B_{\rm cr}} \right)^2 + 0.522 \cdot 10^{-42} \right] {\cal E} \right\}^2 + \bigl(g_{a \gamma \gamma} \, B_T \bigr)^2 \right)^{- \, 1/2}~,
\end{equation}
and the $\gamma \to a$ conversion probability takes the form
\begin{equation}
\label{a18}
P_{\gamma \to a} ({\cal E}, y) = \left(\frac{g_{a \gamma \gamma} \, B_T \, l_{\rm osc} ({\cal E})}{2 \pi} \right)^2 {\rm sin}^2 \left(\frac{\pi y}{l_{\rm osc} ({\cal E})} \right)~, \ \ \ \ \ \ \ \ \ \ \ \ y \leq L_{\rm dom}~.
\end{equation}

Let us next define the {\it low-energy threshold} ${\cal E}_L$ and the {\it high-energy threshold} ${\cal E}_H$ as 
\begin{equation}
\label{eqprop1q}
{\cal E}_L \equiv \frac{|m_a^2 - \omega^2_{\rm pl}|}{2 g_{a \gamma \gamma} \, B_T}~,  
\end{equation}
and 
\begin{equation}
\label{18042018e}
{\cal E}_H \equiv g_{a \gamma \gamma} \, B_T \left[1.42 \cdot 10^{- 4} \, 
\left(\frac{B_T}{B_{\rm cr}} \right)^2 + 0.522 \cdot 10^{- 42} \right]^{- 1}~,
\end{equation} 
respectively.

In order to get a deeper insight into this matter, it is important to note that in $l_{\rm osc} ({\cal E})$  the first term goes like ${\cal E}^{- 1}$, the term with coefficient in square brackets goes like ${\cal E}$ whereas the last term is constant. It looks therefore natural to find out the explicit form of $l_{\rm osc} ({\cal E})$ and $P_{\gamma \to a} ({\cal E}, L_{\rm dom})$ when one of the above three terms dominates over the others. Quite schematically, three different regimes can be naturally singled out. 

\begin{itemize}

\item ${\cal E} < \, {\cal E}_L$ -- This is the {\it low-energy weak mixing regime} where the term $\propto {\cal E}^{- 1}$ dominates. Accordingly, we have
\begin{equation}
l_{\rm osc} ({\cal E}) \simeq \frac{4 \pi \, {\cal E}}{|m_a^2  - \omega_{\rm pl}^2|}~,
\label{18042018c} 
\end{equation}
and
\begin{equation}
P_{\gamma \to a} ({\cal E}, L_{\rm dom}) \simeq \left(\frac{2 \, g_{a \gamma \gamma} \, B_T   \, {\cal E}}{|m_a^2  - \omega_{\rm pl}^2| } \right)^2 {\rm sin}^2 \left(\frac{|m_a^2  - \omega_{\rm pl}^2| \, L_{\rm dom}}{4 \, {\cal E}} \right)~.
\end{equation}
Manifestly, $l_{\rm osc} ({\cal E})$ and $P_{\gamma \to a} ({\cal E}, L_{\rm dom})$ increase with increasing ${\cal E}$, and in addition $P_{\gamma \to a} ({\cal E}, L_{\rm dom})$ exhibits oscillation in ${\cal E}$, which evidently reflects the fact that the {\it individual realizations} of the beam propagation are also oscillating functions of ${\cal E}$. This fact was first noted in~\cite{dmr2008}, brought out explicitly for polarized states in~\cite{wb2012} and generalized to unpolarized states in~\cite{gr2013}. Because $P_{\gamma \to a}  ({\cal E}, L_{\rm dom}) \propto B_T^2$, as $B_T$ decreases the $\gamma \leftrightarrow a$ oscillations at some point become unobservable and the same occurs for decreasing ${\cal E}$.

\item ${\cal E}_L < {\cal E} < {\cal E}_H$ -- This is the intermediate-energy or {\it strong mixing regime} where the ${\cal E} = {\rm constant}$ term dominates. Correspondingly, we get
\begin{equation}
\label{18042018d}
l_{\rm osc} \simeq \frac{2 \pi}{g_{a \gamma \gamma} \, B_T}~,
\end{equation}
and
\begin{equation}
\label{18042018cc} 
P_{\gamma \to a} (L_{\rm dom}) \simeq {\rm sin}^2 \left(\frac{g_{a \gamma \gamma} \, B_T  \, L_{\rm dom}}{2} \right)~.
\end{equation}
Now both $l_{\rm osc}$ and $P_{\gamma \to a} (L_{\rm dom}) $ turn out to be independent both of $m_a$ and of ${\cal E}$, and $P_{\gamma \to a}$ becomes maximal. Note that in this case Eq. (\ref{18042018cc}) justifies our previous statement that the conversion probability depends on the product of the magnetic field times the distance. Moreover, $m_a$ enters ${\cal E}_L$ only.

\item ${\cal E} > {\cal E}_H$ -- This is the {\it high-energy weak mixing regime}, which is in a sense a sort of reversed low-energy weak mixing regime where however the 
QED terms $\propto {\cal E}$ dominate. Hence, we have
\begin{equation}
\label{18042018ee}
l_{\rm osc} ({\cal E}) \simeq \frac{2 \pi}{{\cal E}} \, \left[1.42 \cdot 10^{- 4} \, \left(\frac{B_T}{B_{\rm cr}} \right)^2 + 0.522 \cdot 10^{- 42} \right]^{- 1}~,
\end{equation}
and
\begin{eqnarray}
&\displaystyle  P_{\gamma \to a} ({\cal E}, L_{\rm dom}) \simeq \left(\frac{g_{a \gamma \gamma} \, B_T}{{\cal E}} \right)^2 \left[1.42 \cdot 10^{- 4} \, \left(\frac{B_T}{B_{\rm cr}} \right)^2 + 0.522 \cdot 10^{- 42} \right]^{- 2} \times \nonumber \\ 
&\displaystyle  {\rm sin}^2 \left\{\left(\frac{{\cal E} \, L_{\rm dom}}{2} \right) \left[1.42 \cdot 10^{- 4} \, \left(\frac{B_T}{B_{\rm cr}} \right)^2 + 0.522 \cdot 10^{- 42} \right] \right\}~. \label{18042018f} 
\end{eqnarray}
Evidently now $l_{\rm osc} ({\cal E})$ and $P_{\gamma \to a} ({\cal E}, L_{\rm dom})$ decrease as ${\cal E}$ increases and $P_{\gamma \to a}$ exhibits oscillations in ${\cal E}$ above ${\cal E}_H$: this reflects the fact that the individual realizations of the beam propagation are also oscillating functions of ${\cal E}$. Since $P_{\gamma \to a} \propto {\cal E}^{- 2}$, as ${\cal E}$ increases the $\gamma \leftrightarrow a$ oscillations at some point become unobservable. 

\end{itemize}

\begin{figure}[h]
\centering
\includegraphics[width=.45\textwidth]{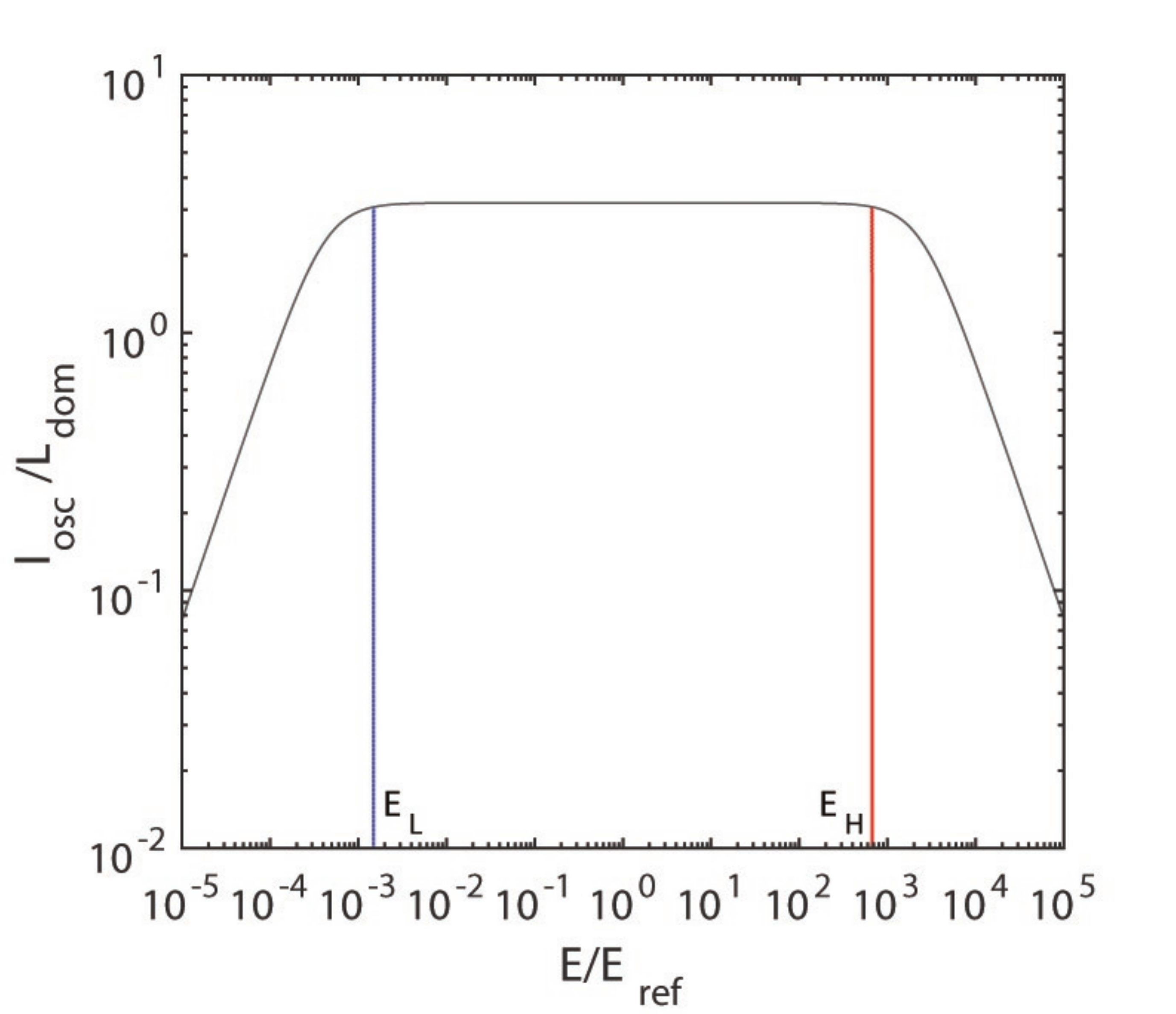}
\caption{\label{LoscComp} Behavior of $l_{\rm osc}/L_{\rm dom}$ versus ${\mathcal E}/{\cal E}_{\rm ref}$ where ${\cal E}_{\rm ref}$ is a reference energy. Also ${\cal E}_L$ and ${\cal E}_H$ are plotted.}
\end{figure}

\begin{figure}[t]
\centering
\includegraphics[width=.50\textwidth]{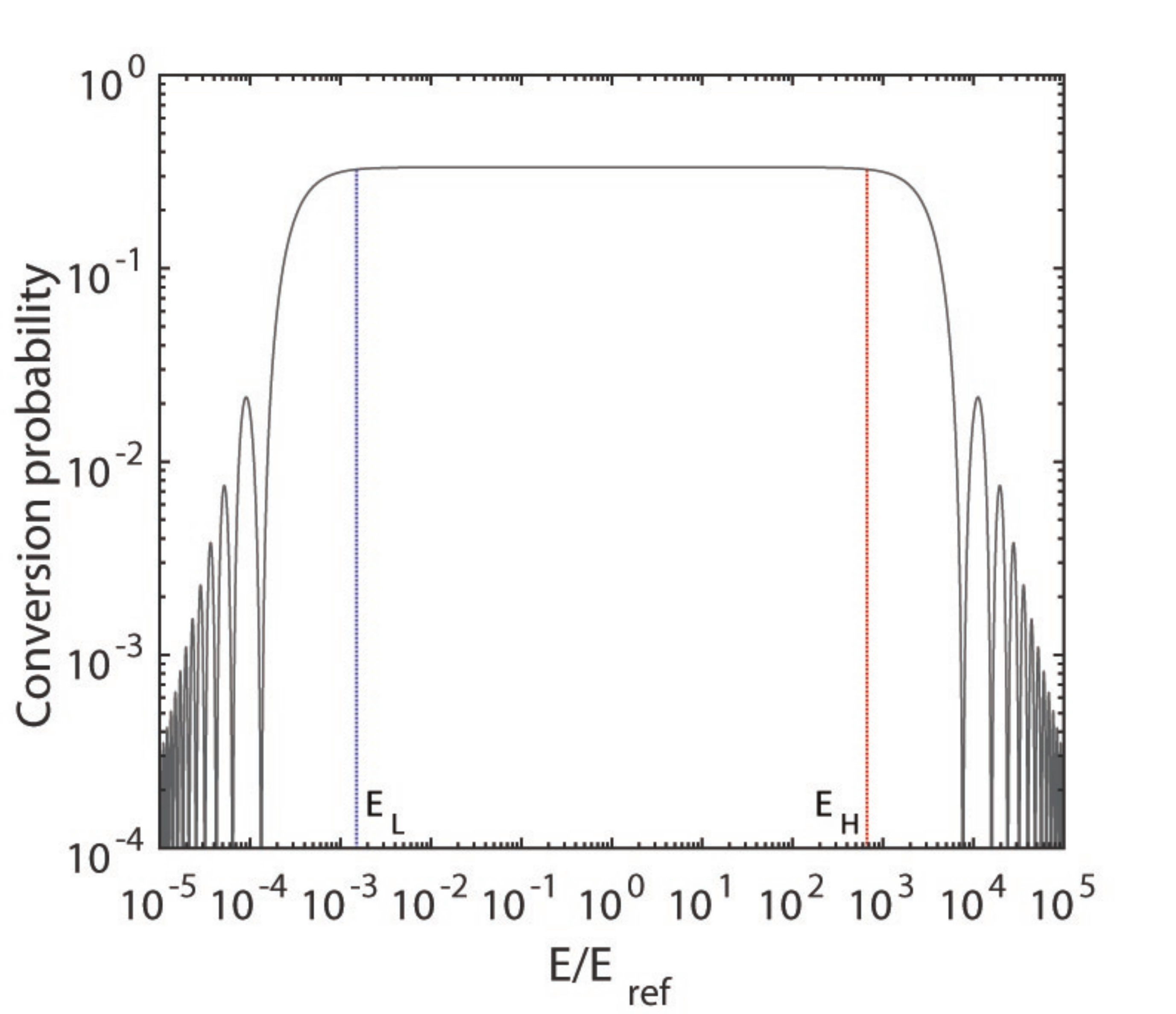}
\caption{\label{ConvPr} Behavior of $P_{\gamma \to a}$ versus ${\mathcal E}/{\cal E}_{\rm ref}$ where ${\cal E}_{\rm ref}$ is a reference energy. Also ${\cal E}_L$ and ${\cal E}_H$ are plotted.}
\end{figure}

We feel that it is very enlightening to have a pictorial view of the behaviour of $l_{\rm osc}$ and $P_{\gamma \to a}  (L_{\rm dom})$ as a function of ${\cal E}$. Because it is convenient to express the energy in dimensionless units, we replace ${\cal E}$ by ${\cal E}/{\cal E}_{\rm ref}$ on the horizontal axis in Figures~\ref{LoscComp} and~\ref{ConvPr} 
(${\cal E}_{\rm ref}$ is an arbitrary reference energy). For the same reason, we replace $l_{\rm osc}$ by $l_{\rm osc}/L_{\rm dom}$ on the vertical axis of Figure~\ref{LoscComp}. Now, Figure~\ref{LoscComp} shows a plot of $l_{\rm osc}$ versus ${\cal E}$, while Figure~\ref{ConvPr} exhibits a plot of $P_{\gamma \to a} (L_{\rm dom})$ versus ${\cal E}$.

\newpage

\section{Implementing our proposal}

Let us consider again the astronomical system ${\cal T}$ defined in Section I, assuming 
that it is magnetized and that the properties of the magnetic field ${\bf B} ( y )$ are the same, namely that ${\bf B} ( y )$ has nearly the more or less the same strength in the region crossed by the beam and that its coherence length is very much smaller than the size of ${\cal T}$. 

As already stressed, our aim is to replace the DLSHE model with the DLSME model.

In order to set up the DLSME model, we employ a {\it linear smoothing} connecting any two adjacent magnetic domains, thereby getting rid of the sharp edges. Even if there is an obvious arbitrariness in our choice, it looks as the most natural one. 

Besides that, given our ignorance of the strength of ${\bf B} ( y )$ in every domain and the previous assumption about the overall strength $B ( y )$, we suppose to average $B ( y )$ over many domains, and next we attribute the resulting value to each domain, so that $B$ -- but not ${\bf B} ( y )$ -- will henceforth be regarded as constant in first approximation. Moreover, in spite of the fact that we suppose that the size of the domains randomly changes within a given range, in order to describe the smoothing procedure as simply as possible, we prefer to avoid taking domains of different size at this stage: it will be done afterwards.

Specifically, our construction of the DLSME model is as follows. Assuming that the photon/ALP beam in question crosses $N$ domains of size $L_{\rm dom}$ during its propagation in ${\cal T}$, let us start by considering the smoothing of the $\phi$ angle. We let $\{y_{D,n}\}_{0 \leq n \leq N}$ be the set of coordinates which defines the beginning ($y_{D,n-1}$) and the end ($y_{D,n}$) of the $n$-th domain. We denote by $\{\phi_n\}_{1 \leq n \leq N}$ the set of angles that ${\bf B}_T ( y )$ forms at the centre of each domain with a fixed fiducial $x$-direction -- equal for all domains -- in the planes $\Pi ( y )$ orthogonal to the beam. And the same convention is assumed for any considered angle. 

We also define the two quantities $y_{0,n}$ and $y_{1,n}$ as
\begin{equation}
\label{x0}
y_{0,n} \equiv y_{D,n} - \frac{\sigma_{\phi}}{2} \bigl(y_{D,n} - y_{D,n-1} \bigr)~, \,\,\,\,\,\,\,\,\,\, (1 \leq n \leq N - 1)~;
\end{equation}
\begin{equation}
\label{x1}
y_{1,n} \equiv y_{D,n} + \frac{\sigma_{\phi}}{2} \bigl(y_{D,n+1} - y_{D,n} \bigr)~, \,\,\,\,\,\,\,\,\,\, (1 \leq n \leq N - 1)~;
\end{equation}
where $\sigma_{\phi} \in [0,1]$ is the {\it smoothing parameter} along the $y$-direction. The interval $[y_{0,n},y_{1,n}]$ is the region where we apply the smoothing procedure, namely where the angle $\phi (y)$ changes smoothly from the value $\phi_{0,n} \equiv \phi_n$ in the $n$-th domain to the value $\phi_{1,n} \equiv \phi_{n+1}$ in the ($n+1$)-th domain. As a consequence, we have $\phi_{0,n+1} \equiv \phi_{1,n}$. Whence $\phi_{0,n+1}\equiv\phi_{1,n}\equiv\phi_{n+1}$. Clearly, for $\sigma_{\phi} = 0$ we get $y_{0,n} =y_{1,n}$, the smoothing region vanishes and we recover the DLSHE model. On the other hand, for $\sigma_{\phi} = 1$ then $y_{0,n}$ becomes the midpoint of the $n$-th domain, and likewise $y_{1,n}$ becomes the midpoint of the $(n+1)$-th domain: in this case the smoothing is maximal, because we never have a constant value of $\phi$ in any domain. The general case is of course represented by a value of $\sigma_{\phi} \in (0,1)$ -- namely somewhat midway -- so that in the central part of a domain the angle is constant ($\phi_{0,n}$) and then it linearly joins the value of the constant angle in the next domain ($\phi_{1,n}$). Therefore, in a generic interval $[y_{1,n-1},y_{1,n}] \, (1 \leq n \leq N-1)$ we have  
\begin{equation}
\label{phi}
\phi (y) = \begin{cases}
\phi_{0,n} = {\rm constant}~, & y \in [y_{1,n-1},y_{0,n}] ~,\\[8pt]
\phi_{0,n} + \dfrac{\phi_{1,n} - \phi_{0,n}}{y_{1,n} - y_{0,n}} \, (y - y_{0,n})~, & y \in [y_{0,n},y_{1,n}]~.
\end{cases} 
\end{equation}
Actually, Eq. (\ref{phi}) defines the behavior of $\phi(y)$ along the whole distance travelled by the photon/ALP beam, provided that two conditions are taken into account.

\begin{itemize}

\item We interpret $y_{1,0} \equiv y_{\rm in}$, where $y_{\rm in}$ is the value of $y$ where the beam enters ${\cal T}$ at the position of the source ${\cal S}$ (origin of the propagation). 

\item Eq. (\ref{phi}) is not valid for $n=N$, which represents the domain where the beam exits from ${\cal T}$: the reason is that there is not a subsequent domain and in this case $\phi(y) = \phi_{0,N} \equiv \phi_N, y \in [y_{1,N-1}, y_{\rm ex}]$, where $y_{\rm ex}$ is the value of $y$ where the beam exits from ${\cal T}$ as interpreted as the position of the observer. 

\end{itemize}

\begin{figure}[h]       
\begin{center}
\includegraphics[width=.90\textwidth]{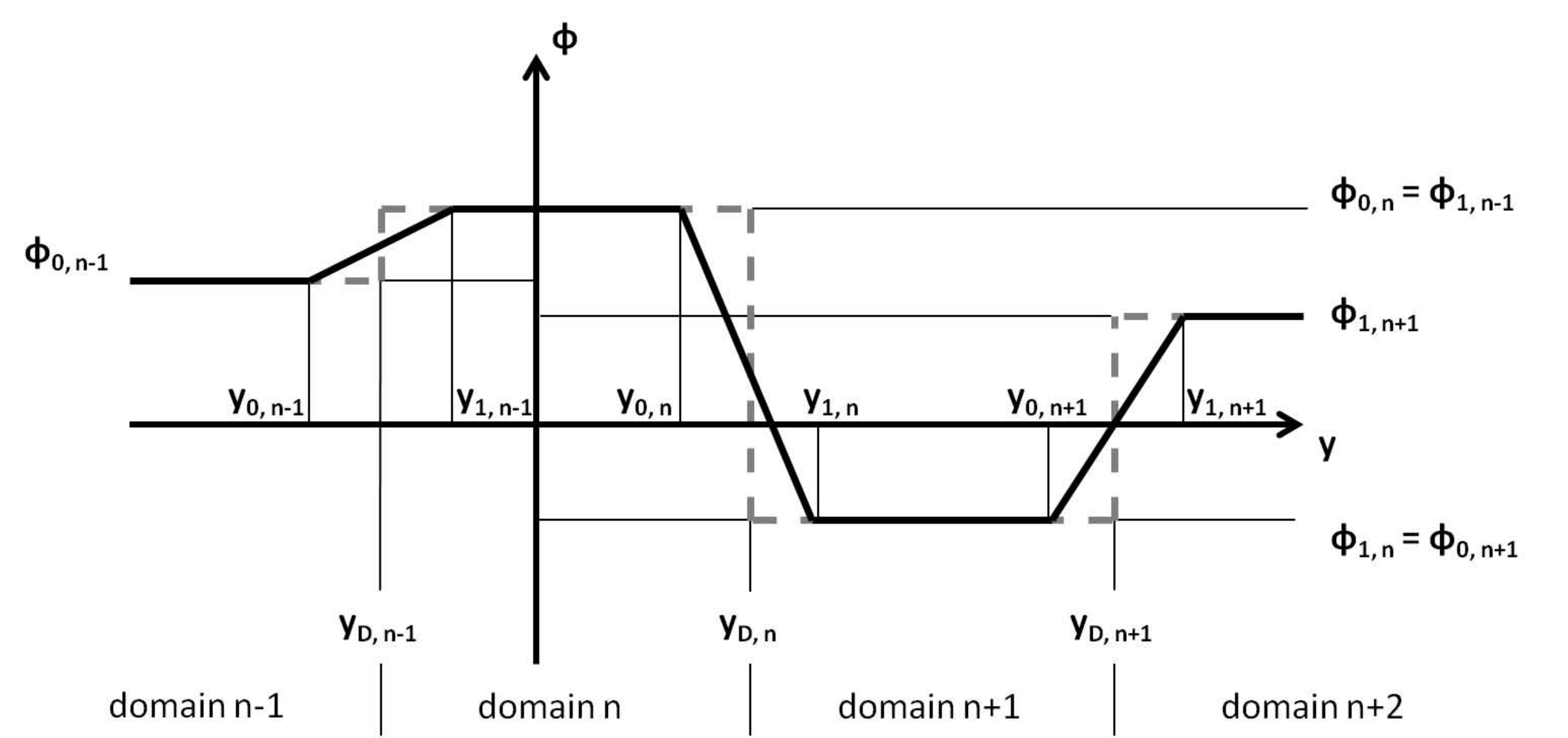}
\end{center}
\caption{\label{linear} 
Behavior of the angle $\phi$ in the $\Pi ( y )$ planes between ${\bf B}_T ( y )$ and the fixed $x$-axis all along the propagation of the beam in the $y$-direction: the solid black line is the new smooth version, while the broken gray line represents its usual jump from one domain to the next. The horizontal solid and broken lines partially overlap. For illustrative simplicity, we have taken the same length for all domains. Note that the source is on the left of the diagram, while the observer is on the right.}
\end{figure}

\noindent This construction is schematically illustrated in Figure~\ref{linear}. Even though the present notation might look somewhat awkward, its advantage is to allow us to get rid of the subindex $n$ in Section VII.

Let us next focus our attention on the smoothing of the $\theta ( y )$ angle, writing -- just as before -- $\{\theta_n\}_{1 \leq n \leq N}$ for the set of angles that ${\bf B}_L ( y )$ forms at the centre of each domain with $y$-direction. Actually, exactly the same procedure as applied above to the $\phi ( y )$ angles can be repeated {\it verbatim} in the present case, and so we do not bore the reader by duplicating the previous discussion. In this case, the smoothing parameter entering the counterpart of Eqs. (\ref{x0}) and (\ref{x1}) is denoted by $\sigma_{\theta}$.

A glance at Eq. (\ref{phi}) -- or even better at Figure~\ref{linear} -- shows that the propagation of the photon/ALP beam in the interval $[y_{1,n-1},y_{0,n}]$ is the usual propagation inside a domain with {\it homogeneous} magnetic field: correspondingly the mixing matrix ${\cal M}$ is constant and the solution of Eq. (\ref{redeqprop}) is presented in Section VI. Accordingly, we denote the corresponding transfer matrix in the $n$-th domain ($1 \leq n \leq N$) by ${\cal U}_{\rm const} \bigl({\cal E}; y_{0, n}, y_{1, n - 1}; \phi_{0, n}, \theta_{0, n} \bigr)$. Instead, the propagation over the interval $[y_{0,n},y_{1,n}]$ is characterized by a lineally varying orientation of ${\bf B} ( y )$: in this case the solution of Eq. (\ref{redeqprop}) is more difficult and it will be derived in Section VII. Accordingly, the transfer matrix in the $n$-th ($1 \leq n \leq N-1$) domain is denoted by ${\cal U}_{\rm var} \bigl({\cal E}; y_{1, n}, y_{0, n}; \phi_{1,n}, \phi_{0,n}; \theta_{1,n}, \theta_{0,n} \bigr)$. 

Altogether, we find that the transfer matrix associated with a single smooth $n$-th domain turns out to be 
\begin{eqnarray}
&\displaystyle {\cal U} \bigl({\cal E}; y_{1, n}, y_{1, n - 1}; \phi_{1,n}, \phi_{1,n-1}; \theta_{1,n}, \theta_{1,n-1} \bigr) = {\cal U}_{\rm var} \bigl({\cal E}; y_{1, n}, y_{0, n}; \phi_{1,n}, \phi_{0,n}; \theta_{1,n}, \theta_{0,n} \bigr) \times \nonumber \\
&\displaystyle {\cal U}_{\rm const} \bigl({\cal E}; y_{0, n}, y_{1, n - 1}; \phi_{0, n}; \theta_{0, n} \bigr)~, \label{mr111117a}
\end{eqnarray}
which is the main result concerning the construction of the DLSME model. 

Next, quantum mechanics implies that the total transfer matrix pertaining to the whole beam propagation through ${\cal T}$ reads
\begin{eqnarray}
&\displaystyle  {\cal U}_{\cal T} \bigl({\cal E}; y_{\rm ex}, y_{\rm in}; \{ \phi_n \}_{1 \le n \le N}, \{ \theta_n \}_{1 \le n \le N} \bigr) = {\cal U}_{\rm const} \bigl({\cal E}; y_{\rm ex}, y_{1, N - 1}; \phi_{0, N}, \theta_{0, N} \bigr) \times   \label{utot} \\
&\displaystyle \prod_{n=1}^{N-1} {\cal U}_{\rm var} \bigl({\cal E}; y_{1, n}, y_{0, n}; \phi_{1,n}, \phi_{0,n}; \theta_{1,n}, \theta_{0,n} \bigr) \ {\cal U}_{\rm const} \bigl({\cal E}; y_{0, n}, y_{1, n - 1}; \phi_{0, n}, \theta_{0, n} \bigr)~. \nonumber
\end{eqnarray}

A notational remark is now important. While in the transfer matrix as taken alone the angles  vary with $y$, since such a variation smoothly interpolates between the constant values of the angles in adjacent domains, in the transfer matrix across the whole ${\cal T}$ the {\it constant} values of the angles in all domains {\it completely fix} $\phi (y)$ and $\theta (y)$, as it is clear from Eq. (\ref{phi}): this fact explains the notation employed in Eq. (\ref{utot}). Just like in the previous Section -- using Eq. (\ref{unpprob}) with $\rho_0 \equiv \rho_{\rm in}$, $\rho \equiv \rho_{\rm ex}$ as applied to ${\cal T}$ -- the associated photon survival probability for a single realization of the beam propagation process is 
\begin{eqnarray}
&\displaystyle P^{\rm ALP}_{\gamma \to \gamma} \bigl({\cal E}; y_{\rm ex}, \rho_{\rm ex}; y_{\rm in}, \rho_{\rm in}; \, \{\phi_n \}_{1 \le n \le N}, \{\theta_n \}_{1 \le n \le N} \bigr) = \label{mr111117b} \\
&\displaystyle {\rm Tr} \left[\rho_{\rm ex} \, {\cal U}_{\cal T}  
\bigl({\cal E}; y_{\rm ex}, y_{\rm in}; \{ \phi_n \}_{1 \le n \le N}, \{ \theta_n \}_{1 \le n \le N} \bigr) \, \rho_{\rm in} \, {\cal U}_{\cal T}^{\dagger} \bigl({\cal E}; y_{\rm ex}, y_{\rm in}; \{ \phi_n \}_{1 \leq n \leq N}, \{ \theta_n \}_{1 \leq n \leq N} \bigr) \right]~, \nonumber
\end{eqnarray} 

We recall that since only a single realization of the beam propagation is actually observed, Eq. (\ref{mr111117b}) is indeed the key-quantity.

\subsection{Three-dimensional numerical solution} 

\begin{figure}[h]
\centering
\includegraphics[width=1.10\textwidth]{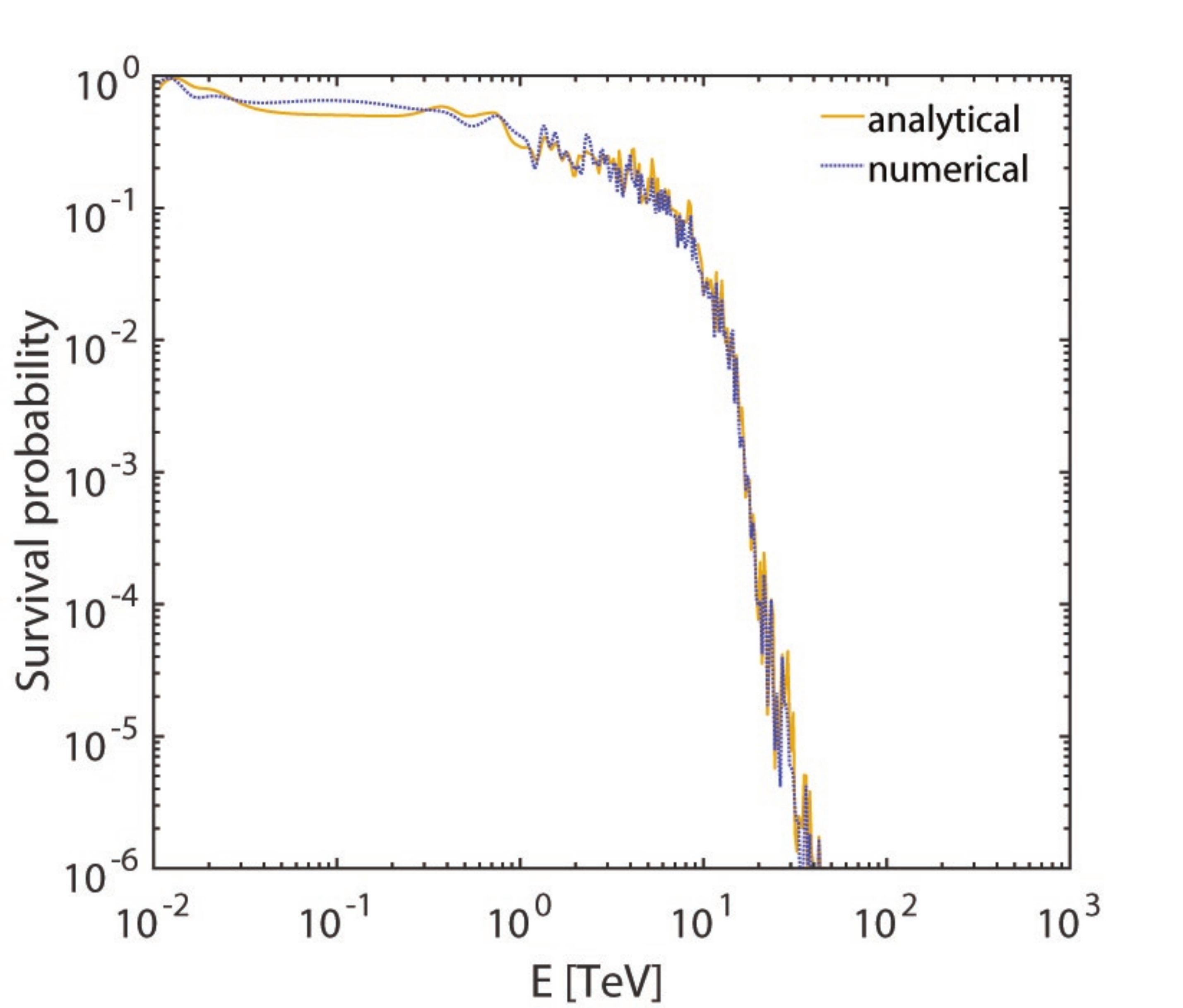}
\caption{\label{confronto} Comparison between the three-dimensional numerical solution and the two-dimensional analytical solution for the above choice of the parameters.}
\end{figure}

In general, the beam propagation equation (\ref{redeqprop}) has the form
\begin{equation}
\label{07032018a}
\left[ i \frac{d}{d y} +  \left(
\begin{array}{ccc}
w ({\cal E}, y ) & 0 & v ( y ) \, s( y ) \\
0 & w ({\cal E}, y ) & v ( y ) \, c( y ) \\
v ( y ) \, s( y ) & v ( y ) \, c( y ) & u ({\cal E})
\end{array}
\right) \right] \, \psi (y) = 0~,
\end{equation}

Unfortunately, such an equation looks very difficult to solve analytically, and so we resort to a numerical treatment. As a case study -- we shall come back to this topic in a systematic fashion elsewhere~\cite{paper2} -- we take ${\cal T}$ to be the extragalactic space out to redshift $z = 0.1$ where a blazar is located and supposed to emit in the VHE band. Actually, ${\cal T}$ is just the kind of system that absorbs photons. It is well in fact known that the infrared/optical/ultraviolet sky is dominated by the {\it extragalactic background light} (EBL), which is the total light emitted by all galaxies during their cosmic evolution (for a review, see~\cite{dwek}). So, whenever a VHE photon  scatters off an EBL photon there is a chance to produce an $e^+ e^-$ pair, which causes a dimming of the blazar. We use the Franceschini and Rodighiero EBL model~\cite{franceschinirodighiero}. Hence, $w ({\cal E}, y )$ takes EBL and photon dispersion on the CMB into account (the plasma frequency can safely be neglected). We also recall that $v (y) \, s ( y )$ and $v (y) \, c( y )$ describe the photon-ALP coupling (we take $B_T/({\rm nG}) = g_{a \gamma \gamma} \, 10^{11} \, {\rm GeV}$ for definiteness) and for the ALP mass we choose for instance $m_a = 10^{- 10} \, {\rm eV}$. As far as the probability density of the domain size is concerned, we take a power law distribution function $\propto L_{\rm dom}^{- 1.2}$ inside the range $0.2 \, {\rm Mpc} - 10 \, {\rm Mpc}$, which entails that $\langle L_{\rm dom} \rangle = 2 \, {\rm Mpc}$, which is in agreement with the present bounds~\cite{durrerneronov}. Because the emitted beam is generally supposed to be unpolarized and the polarization cannot be measured in the VHE band, Eq. (\ref{mr111117b}) presently reads
\begin{eqnarray}
&\displaystyle P^{\rm ALP}_{\gamma \to \gamma, {\rm unp}} \bigl({\cal E}; y_{\rm ex}, \rho_x, \rho_z; y_{\rm in}, \rho_{\rm unp}; \, \{\phi_n \}_{1 \le n \le N}, \{\theta_n \}_{1 \le n \le N} \bigr) = \label{0903208c} \\
&\displaystyle \sum_{i = x,z} {\rm Tr} \left[\rho_i \, {\cal U}_{\cal T}  
\bigl({\cal E}; y_{\rm ex}, y_{\rm in}; \{ \phi_n \}_{1 \le n \le N}, \{ \theta_n \}_{1 \le n \le N} \bigr) \, \rho_{\rm unp} \, {\cal U}_{\cal T}^{\dagger} \bigl({\cal E}; y_{\rm ex}, y_{\rm in}; \{ \phi_n \}_{1 \leq n \leq N}, \{ \theta_n \}_{1 \leq n \leq N} \bigr) \right]~. \nonumber
\end{eqnarray}
with 
\begin{equation}
\label{sm3a}
\rho_x \equiv
\left(
\begin{array}{ccc}
1 & 0 & 0 \\
0 & 0 & 0 \\
0 & 0 & 0
\end{array}
 \right)~,\,\,\,\,\,\,\,\,\,\,
\rho_z \equiv
\left(
\begin{array}{ccc}
0 & 0 & 0 \\
0 & 1 & 0 \\
0 & 0 & 0
\end{array}
 \right)~,\,\,\,\,\,\,\,\,\,\,
\rho_{\rm unp} \equiv
\frac{1}{2}\left(
\begin{array}{ccc}
1 & 0 & 0 \\
0 & 1 & 0 \\
0 & 0 & 0
\end{array}
\right)~.
\end{equation}

The result is shown by the dotted line in Figure~\ref{confronto}, where we take $\sigma_{\phi} = \sigma_{\theta} = 0.2$ for the smoothing parameters. We emphasize that we have checked that the same conclusion remains true also for different choices of the parameters.

\subsection{Two-dimensional analytical solution} 

Yet, because the longitudinal component ${\bf B}_L ( y )$ of ${\bf B} ( y )$ does not couple to ALPs we are naturally led to guess that what really matters is what happens in the planes $\Pi ( y )$. Owing to our ignorance about ${\bf B} ( y )$, we proceed as we did at the beginning of this Section: we now take for consistency on average $B_T = \sqrt{2/3} \, B$, over many domains, which is next attributed to every domain. As a consequence, it looks very remarkable that Eqs. (\ref{07032018a}) become amenable of an exact -- even if non-trivial -- treatment since the quantities $w ({\cal E})$ and $v$ in the beam propagation equation (\ref{07032018a}) turn out to be constant with respect to $y$. 

Manifestly -- as Figure~\ref{confronto} shows -- there is basically no difference between the two cases, namely the three-dimensional numerical solution and two-dimensional analytical solution to be derived later. Given this situation, we prefer to deal with the analytical two-dimensional solution both in order to have a complete control of what goes on and mainly because it leads to a {\it dramatic reduction} in the computation time in the numerical applications. As a consequence, the $\theta$ angle disappears from the game and consequently we shall denote the smoothing parameter simply as $\sigma$.

Incidentally, the agreement between the two cases flags the fact that the true physics of the considered problem is confined inside the planes $\Pi  ( y )$  indeed in agreement with our guess. Therefore, now Eq. (\ref{utot}) reduces to 
\begin{eqnarray}
&\displaystyle  {\cal U}_{\cal T} \bigl({\cal E}; y_{\rm ex}, y_{\rm in}; 
\{ \phi_n \}_{1 \le n \le N} \bigr) = {\cal U}_{\rm const} \bigl({\cal E}; y_{\rm ex}, y_{1, N - 1}; \phi_{0, N} \bigr) \times   \label{0105201b} \\
&\displaystyle \prod_{n=1}^{N-1} {\cal U}_{\rm var} \bigl({\cal E}; y_{1, n}, y_{0, n}; \phi_{1,n}, \phi_{0,n} \bigr) \ {\cal U}_{\rm const} \bigl({\cal E}; y_{0, n}, y_{1, n - 1}; \phi_{0, n} \bigr)~, \nonumber
\end{eqnarray}
and so the photon survival probability along a single realization of the unpolarized beam becomes 
\begin{eqnarray}
&\displaystyle P^{\rm ALP}_{\gamma \to \gamma, {\rm unp}} \bigl({\cal E}; y_{\rm ex}, \rho_x, \rho_z; y_{\rm in}, \rho_{\rm unp}; \, \{\phi_n \}_{1 \le n \le N} \bigr) = \label{12042018a} \\
&\displaystyle \sum_{i = x,z} {\rm Tr} \left[\rho_i \, {\cal U}_{\cal T}  
\bigl({\cal E}; y_{\rm ex}, y_{\rm in}; \{ \phi_n \}_{1 \le n \le N} \bigr) \, \rho_{\rm unp} \, {\cal U}_{\cal T}^{\dagger} \bigl({\cal E}; y_{\rm ex}, y_{\rm in}; \{ \phi_n \}_{1 \leq n \leq N} \bigr) \right]~, \nonumber
\end{eqnarray}
with $\rho_x$, $\rho_z$, $\rho_{\rm unp}$ defined by Eq. (\ref{sm3a}).

Finally, the propagation inside the constant-angle region is provided in Section VI, whereas the propagation inside a varying-orientation magnetic field is reported in Section VII.

\section{Propagation inside the constant angle region}

Let us begin by evaluating the transfer matrix within a {\it single} $n$-th domain inside its regions where ${\bf B}$ is homogeneous and the quantities $w ({\cal E})$, $v$, $c$, $s$ in Eq. (\ref{07032018a}) are all constant with respect to $y$; for notational simplicity the 
${\cal E}$-dependence of $w$ and $u$ will henceforth not explicitly exhibited. For the same reason, we drop all $n$-subindexes in ${\cal U}_{\rm const} ({\cal E}; y, y_0; \phi)$.

\subsection{General treatment}

We start by diagonalizing the corresponding mixing matrix (\ref{sm3}), thereby finding its  eigenvalues
\begin{equation}
\label{a91212a1app}
{\lambda}_{1} = w~,
\end{equation}
\begin{equation}
\label{a91212a2app}
{\lambda}_{2} = \frac{1}{2} \left( w + u - \sqrt{\left(w - u \right)^2 + 4 \, v^2} \right)~, 
\end{equation}
\begin{equation}
\label{a91212a3app}
{\lambda}_{3} = \frac{1}{2} \left( w + u + \sqrt{\left(w - u \right)^2 + 4 \, v^2} \right)~, 
\end{equation}
and we recall that $u$, $v$, $w$ are given by Eqs. (\ref{mr15112017b}), (\ref{mr15112017c}), (\ref{mr15112017a}), respectively. It is straightforward to check that the corresponding eigenvectors can be taken to be
\begin{equation}
\label{k3w1Aappa1}
X_1 = \left(
\begin{array}{c}
{\rm cos} \, \phi \\
- \, {\rm sin} \, \phi \\
0 \\
\end{array}
\right)~,
\end{equation}
\begin{equation}
\label{k3w1Aappa2}
X_2 = \left(
\begin{array}{c}
v \, {\rm sin} \, \phi \\
v \, {\rm cos} \, \phi \\
{\lambda}_2 - w \\
\end{array}
\right)~,
\end{equation}
\begin{equation}
\label{k3w1Aappa3}
X_3 = \left(
\begin{array}{c}
v \, {\rm sin} \, \phi \\
v \, {\rm cos} \, \phi \\
{\lambda}_3 - w  \\
\end{array}
\right)~.
\end{equation}
Correspondingly, any solution of Eq. (\ref{redeqprop}) can be represented in the form
\begin{equation}
\label{a91212a1appQ}
\psi (y) = c_1 \, X_1 \, e^{i {\lambda}_1 \, \left(y - y_0 \right)} + c_2 \, X_2 \, e^{i {\lambda}_2 \, \left(y - y_0 \right)} + c_3 \, X_3 \, 
e^{i {\lambda}_3 \, 
\left(y - y_0 \right)}~,
\end{equation}
where $c_1$, $c_2$, $c_3$ and $y_0$ are arbitrary constants. As a consequence, the solution with initial condition 
\begin{equation}
\label{k3w1appW}
\psi (y_0) \equiv \left(
\begin{array}{c}
\gamma_1 (y_0) \\
\gamma_2 (y_0) \\
a (y_0) \\
\end{array}
\right)
\end{equation}
emerges from Eq. (\ref{a91212a1appQ}) for
\begin{equation}
\label{a91212a1appq1}
c_{1} = \gamma_1 (y_0) \, {\rm cos} \, \phi - \gamma_2 (y_0) \, {\rm sin} \, \phi~,
\end{equation}
\begin{equation}
\label{a91212a1appq2}
c_{2} = \frac{{\lambda}_3 - w}{v ({\lambda}_3 - {\lambda}_2 )} \, {\rm sin} \, \phi \, \gamma_1 (y_0) + \frac{{\lambda}_3 - w}{v ({\lambda}_3 - {\lambda}_2 )} \, {\rm cos} \, \phi \, \gamma_2 (y_0) - \frac{1}{{\lambda}_3 - {\lambda}_2} \, a(y_0)~,
\end{equation}
\begin{equation}
\label{a91212a1appQ3}
c_{3} = - \frac{{\lambda}_2 - w}{v ({\lambda}_3 - {\lambda}_2 )} \, {\rm sin} \, \phi \, \gamma_1 (y_0) - \frac{{\lambda}_2 - w}{v ({\lambda}_3 - {\lambda}_2 )} \, {\rm cos} \, \phi \, \gamma_2 (y_0) + \frac{1}{{\lambda}_3 - {\lambda}_2} \, a(y_0)~.
\end{equation}
It is a simple exercise to recast the considered solution into the form
\begin{equation}
\label{k3lasqapp}
\psi (y) = {\cal U}_{\rm const} ({\cal E}; y, y_0; \phi) \, \psi (y_0) 
\end{equation}
with
\begin{equation}
\label{mravvq2abcapp}
{\cal U}_{\rm const} ({\cal E}; y, y_0; \phi) = e^{i {\lambda}_1 (y - y_0)} \, T_1 (\phi) + e^{i {\lambda}_2 (y - y_0)} \, T_2 (\phi) + e^{i {\lambda}_3 (y - y_0)} \, T_3 (\phi)~, 
\end{equation}
where we have set
\begin{equation}
\label{mravvq1app}
T_1 (\phi) \equiv
\left(
\begin{array}{ccc}
{\rm cos}^2 \phi & - \, {\rm sin} \, \phi \, {\rm cos} \, \phi& 0 \\
- \ {\rm sin} \, \phi \, {\rm cos} \, \phi & {\rm sin}^2 \phi & 0 \\
0 & 0 & 0
\end{array}
\right)~,
\end{equation}
\begin{equation} 
\label{mravvq2app}
T_2 (\phi) \equiv
\left(
\begin{array}{ccc}
\frac{{\lambda}_3 - w}{{\lambda}_3 - {\lambda}_2} \, {\rm sin}^2 \phi & \frac{{\lambda}_3 - w}{{\lambda}_3 - {\lambda}_2} \, {\rm sin} \, \phi \, {\rm cos} \, \phi &  - \, \frac{v}{{\lambda}_3 - {\lambda}_2} \, {\rm sin} \, \phi \\ 
\frac{{\lambda}_3 - w}{{\lambda}_3 - {\lambda}_2} \, {\rm sin} \, \phi \, {\rm cos} \, \phi & \frac{{\lambda}_3 - w}{{\lambda}_3 - \, {\lambda}_2} \, {\rm cos}^2 \phi & - \, \frac{v}{{\lambda}_3 - {\lambda}_2} \, {\rm cos} \, \phi \\
 - \, \frac{v}{{\lambda}_3 - {\lambda}_2} \, {\rm sin} \, \phi &  - \, \frac{v}{{\lambda}_3 - {\lambda}_2} \, {\rm cos} \, \phi &  - \, \frac{{\lambda}_2 - w}{{\lambda}_3 - {\lambda}_2}
\end{array}
\right)~,
\end{equation} 
\begin{equation} 
\label{mravvq2app}
T_3 (\phi) \equiv
\left(
\begin{array}{ccc}
- \, \frac{{\lambda}_2 - w}{{\lambda}_3 - {\lambda}_2} \, {\rm sin}^2 \phi & - \, \frac{{\lambda}_2 - w}{{\lambda}_3 - {\lambda}_2} \, {\rm sin} \, \phi \, {\rm cos} \, \phi & \frac{v}{{\lambda}_3 - {\lambda}_2} \, {\rm sin} \, \phi \\ 
- \, \frac{{\lambda}_2 - w}{{\lambda}_3 - {\lambda}_2} \, {\rm sin} \, \phi \, {\rm cos} \, \phi & - \, \frac{{\lambda}_2 - w}{{\lambda}_3 - \, {\lambda}_2} \, {\rm cos}^2 \phi & \frac{v}{{\lambda}_3 - \, {\lambda}_2} \, {\rm cos} \, \phi \\ 
\frac{v}{{\lambda}_3 - {\lambda}_2} \, {\rm sin} \, \phi & \frac{v}{{\lambda}_3 - {\lambda}_2} \, {\rm cos} \, \phi & \frac{{\lambda}_3 - w}{{\lambda}_3 - {\lambda}_2}
\end{array}
\right)~.
\end{equation} 

\subsection{Recovering the DLSHE model}

As a by-product, we show how the DLSHE model emerges as a particular case within the present context. Actually, what is usually done is to treat the two angles $\phi$ and $\theta$ on a different footing: while $\phi$ is retained, one averages over $\theta$ right at the beginning, so that the relation $B_T = B \, {\rm sin} \, \theta$ gets replaced by $B_T = 
\sqrt{2/3} \, B$.

Therefore, we proceed to address the propagation over the {\it whole} ${\cal T}$. To this end, it is convenient to slightly change our notation, writing ${\cal U}_{\rm const} ({\cal E}; y, y_0; \phi) \equiv {\cal U}_{\rm const} \bigl({\cal E}; y_n, y_{n - 1}; \phi_n)$, denoting by 
$y_{\rm in} \equiv y_0$ and $y_{\rm ex} \equiv y_N$ the values of $y$ where the beam enters ${\cal T}$ and exits from ${\cal T}$, respectively (in agreement with Section I). Just as before, the transfer matrix across ${\cal T}$ for a single realization of the beam propagation process -- corresponding to a single choice of all $\{\phi_n\}_{1 \le n \le N}$ -- is given by
\begin{equation}
\label{mr051117a}
{\cal U}_{\rm const} \bigl({\cal E}; y_{\rm ex}, y_{\rm in}; \{\phi_n\}_{1 \le n \le N} \bigr) = \prod_{n=1}^{N} {\cal U}_{\rm const} \bigl({\cal E}; y_n, y_{n - 1}; \phi_n \bigr)~.
\end{equation}
Hence, thanks to Eq. (\ref{unpprob}) -- with $\rho_0 \equiv \rho_{\rm in}$, $\rho \equiv \rho_{\rm ex}$ and applied to ${\cal T}$ -- the associated photon survival probability reads
\begin{eqnarray}
 \label{mr051017b}
&\displaystyle P^{\rm ALP}_{\gamma \to \gamma} \bigl({\cal E};y_{\rm ex}, \rho_{\rm ex}; y_{\rm in}, \rho_{\rm in}; \{\phi_n\}_{1 \le n \le N} \bigr) = \\
&\displaystyle {\rm Tr} \left[\rho_{\rm ex} \, {\cal U}_{\rm const} \bigl({\cal E}; y_{\rm ex}, y_{\rm in}; \{\phi_n \}_{1 \le n \le N} \bigl) \, \rho_{\rm in} \, {\cal U}^{\dagger}_{\rm const} 
\bigl({\cal E}; y_{\rm ex}, y_{\rm in}; \{\phi_n \}_{1 \le n \le N} \bigr) \right]~. \nonumber
\end{eqnarray}

In view of the polarimetric satellite mission IXPE (Imaging X-ray Polarimetry Explorer)~\cite{ixpe} operated by NASA, the proposed polarimetric satellite mission XIPE (X-ray Imaging Polarimetry Explorer)~\cite{xipe}, the conventional/polarimetric 
e-ASTROGRAM mission~\cite{eastrogam1,eastrogam2} and the conventional/polarimetric AMEGO (All-Sky Medium-Energy Gamma-Ray Observatory)~\cite{amego}, we prefer to leave the initial and final polarizations unspecified. However, if the beam is unpolarized and/or the polarization is not measured Eq. (\ref{mr051017b}) reduces to 
\begin{eqnarray}
&\displaystyle P^{\rm ALP, {\rm av}}_{\gamma \to \gamma, {\rm unp}} \bigl({\cal E}; y_{\rm ex}, \rho_x, \rho_z; y_{\rm in}, \rho_{\rm unp} \bigr) = \label{31052018a} \\
&\displaystyle \sum_{i = x,z} \left\langle {\rm Tr} \left[\rho_i \, {\cal U}_{\cal T}  
\bigl({\cal E}; y_{\rm ex}, y_{\rm in}; \{ \phi_n \}_{1 \le n \le N} \bigr) \, \rho_{\rm unp} \, {\cal U}_{\cal T}^{\dagger} \bigl({\cal E}; y_{\rm ex}, y_{\rm in}; \{ \phi_n \}_{1 \leq n \leq N} \bigr) \right] \right\rangle_{\{\phi_n\}_{1 \le n \le N}}~, \nonumber
\end{eqnarray}
where $\rho_x$, $\rho_z$, $\rho_{\rm unp}$ are defined by Eq. (\ref{sm3a}).

\section{Propagation inside a varying-orientation magnetic field}

Let us next evaluate the transfer matrix within a {\it single} $n$-th domain inside its range where $\phi (y)$ varies linearly with $y$. Because it applies to any domain, we simply write it as ${\cal U}_{{\rm var}} \bigl({\cal E}; y, y_{0,n}; \phi (y) \bigr)$ -- with $y \in[y_{0,n}, y_{1,n}]$ and dropping any other subindex $n$ for notational simplicity -- which is associated with the reduced Schr\"odinger-like equation (\ref{redeqprop}) within the interval in question. 

Accordingly, we insert $\phi (y) = \phi_{0,n} + (\phi_{1,n} - \, \phi_{0,n}) (y \, - y_{0,n})/(y_{1,n} - \, y_{0,n})$ from Eq. (\ref{phi}) into the mixing matrix (\ref{sm3}). As in the previous Section, the ${\cal E}$-dependence of the quantities $w$ and $u$ is not explicitly exhibited and we recall that now they turn out to be independent of $y$, as discussed in Section V.B.

Consequently, Eq. (\ref{redeqprop}) splits into the three coupled equations

\begin{equation}  
\label{sm6}
\begin{cases}
\dfrac{d}{dy} \, \gamma_1 ( y ) = i w \, \gamma_1 ( y ) + i v \, s (y) \, a( y )~,\\[8pt]
\dfrac{d}{dy} \, \gamma_2 ( y ) = i w \, \gamma_2 ( y ) + i v \, c (y) \, a( y )~,\\[8pt]
\dfrac{d}{dy} \, a( y ) = i v \bigl[s (y) \, \gamma_1 ( y ) + c (y) \, \gamma_2 ( y ) \bigr] + i u \, a(y)~,
\end{cases}
\end{equation}
to be solved with the initial conditions $\gamma_1 ( y_0 ) = \gamma_{1,0}$, $\gamma_2 (y_0) = \gamma_{2,0}$, $a(y_0) = a_0$ and we recall that $u$, $v$, $w$ are given by Eqs. (\ref{mr15112017b}), (\ref{mr15112017c}), (\ref{mr15112017a}), respectively.

It proves very useful to perform the change of variable
\begin{equation}
\label{sm7}
y \to \zeta \equiv \frac{\phi_1 - \, \phi_0}{y_1 - \, y_0}(y - \, y_0) = k^{- 1} (y - y_0)~,
\end{equation}
where we have set 
\begin{equation}
\label{sm7a}
k^{-1} \equiv \frac{d \zeta}{d y} = \frac{\phi_1 - \, \phi_0}{y_1 - \,y_0}~,  
\end{equation}
which induces the transformations $s (y) \to {\rm sin} (\phi_0 + \zeta)$, 
$c (y) \to {\rm cos} (\phi_0 + \zeta)$ along with 
$\gamma_i ( y ) \to \gamma_i \bigl(y_0 + \frac{y_1 - \, y_0}{\phi_1 - \, \phi_0} \zeta \bigr)$ ($i=1,2$) and $a(y) \to a \bigl(y_0 + \frac{y_1 - \, y_0}{\phi_1 - \, \phi_0} \zeta \bigr)$. Using Eq. (\ref{sm7}) and setting further
\begin{equation}
\label{sm8q}
\overline{\gamma}_i ( \zeta ) \equiv \gamma_i \left(y_0 + \frac{y_1 - \, y_0}{\phi_1 - \phi_0} \zeta \right)~, \ \ \ (i = 1, 2)~; 
\end{equation}
\begin{equation}
\label{sm8w}
\overline{a} ( \zeta ) \equiv a \left(y_0 + \frac{y_1 - \, y_0}{\phi_1 - \, \phi_0} \zeta \right)~,
\end{equation}
we get
\begin{equation}
\label{sm8aq}
\frac{d}{dy} \, \gamma_i \left(y_0 + \frac{y_1 - y_0}{\phi_1 - \phi_0} \zeta \right) = k^{-1} \frac{d}{d\zeta} \, \overline{\gamma}_i(\zeta)~, \ \ \ (i = 1, 2)~; 
\end{equation}
\begin{equation}
\label{sm8aw}
\frac{d}{dy} a \left(y_0 + \frac{y_1 - y_0}{\phi_1 - \phi_0} \zeta \right) = k^{-1} \frac{d}{d\zeta} \, \overline{a} (\zeta)~.
\end{equation}
Hence, the system (\ref{sm6}) can be rewritten as
\begin{equation}
\label{sm9}
\begin{cases}
\dfrac{d}{d\zeta} \, \overline{\gamma}_1 (\zeta) = i k w \, \overline{\gamma}_1 (\zeta) + i k v \ {\rm sin} (\phi_0 + \zeta) \, \overline{a}(\zeta)~,\\[8pt]
\dfrac{d}{d\zeta} \, \overline{\gamma}_2(\zeta) = i k w \, \overline{\gamma}_2 (\zeta) + i k v \ {\rm cos} (\phi_0 + \zeta) \,\overline{a}(\zeta)~,\\[8pt]
\dfrac{d}{d\zeta} \, \overline{a} (\zeta) = i k v \bigl[{\rm sin} (\phi_0 + \zeta) \, \overline{\gamma}_1 (\zeta) + {\rm cos} (\phi_0 + \zeta) \,\overline{\gamma}_2(\zeta) \bigr] + i k u \,\overline{a} (\zeta)~,
\end{cases}
\end{equation}
and the initial conditions become $\overline{\gamma}_1 ( 0 ) = \gamma_{1,0}$, $\overline{\gamma}_2 ( 0 ) = \gamma_{2,0}$, $\overline{a} ( 0 ) = a_0$.

The system (\ref{sm9}) is now in a form suitable to perform a Laplace transform $\mathcal L$ provided that all functions in Eqs.  (\ref{sm9}) are $\mathcal L$-transformable. The causal unilateral Laplace transform $\mathcal L_+$ applies to functions 
$f(\zeta)$ defined for all real numbers $\zeta \ge 0$ provided that $f(\zeta)$ is locally integrable in the interval $[0,+\infty)$. Correspondingly, the Laplace transform of $f ( \zeta )$ is
\begin{equation}
\label{sm10}
F(s) = {\mathcal L}_+\{f\}(s)=\int_0^{+\infty} e^{-s\zeta} \, f(\zeta) \, d \zeta~,
\end{equation}
where $s$ is a complex variable. It is clear that all functions in the system (\ref{sm9}) are locally integrable: indeed, in order for the $\mathcal L$-transform to exist, it is enough that all functions grow less than exponentially and this is certainly true owing to their physical meaning. In addition, all functions are defined in the bounded interval $[y_0, y_1]$. However, a problem arises concerning the values taken by the $\zeta$ variable. Since we have $y \in[y_0,y_1]$, we see from Eq. (\ref{sm7}) that $\zeta \in[0,\phi_1 
- \phi_0]$ for $\phi_1 > \phi_0$ and $\zeta \in[\phi_1-\phi_0,0]$ for $\phi_1 < \phi_0$. While there is no problem in the former case, clearly in the latter case no ${\mathcal L}_+$-transform exists. A way out of this difficulty amounts to employ the anti-causal unilateral Laplace transform $\mathcal L_-$ which is defined for a function $f(\zeta)$ defined for all real numbers $\zeta \le 0$ and which is locally integrable in the interval $(-\infty,0]$. In this case the Laplace transform of $f ( \zeta )$ is
\begin{equation}
\label{sm11}
F( s ) = {\mathcal L}_- \{f \} ( s ) = \int_{- \infty}^0 e^{-s \zeta} \, f( \zeta ) \, d \zeta~,
\end{equation}
where $s$ is again a complex variable. Although we have to deal with the two cases separately in order to solve the system (\ref{sm9}), we will discover that the two solutions are identical, thereby avoiding to worry about the sign of $\phi_1-\phi_0$. We focus here our attention on the case $\phi_1-\phi_0 >0$, since the other is completely similar (apart from two changes that will be mentioned in due place). 

Thanks to Eq. (\ref{sm10}) and using a similar notation -- employing capital letters for the Laplace transform of original functions (dropping overline for simplicity) -- the system (\ref{sm9}) can be $\mathcal L_+$-transformed as  
\begin{equation}
\label{sm12}
\begin{cases}
\Gamma_1(s)=\dfrac{\gamma_{1,0}}{s - i k w} + \dfrac{k v}{2 (s - i k w)} \left(e^{i \phi_0} A( s - i) - \, e^{- i \phi_0} A( s + i) \right)~, \\[8pt]
\Gamma_2(s)=\dfrac{\gamma_{2,0}}{s - i k w} + \dfrac{i k v}{2 (s - i k w)} \left(e^{i \phi_0} A( s - \, i) + e^{- i \phi_0} A(s + i) \right)~, \\[8pt]
s \, A(s) - a_0=\dfrac{k v}{2} \left[e^{i \phi_0} \Gamma_1(s - i) - e^{-i \phi_0} \Gamma_1(s + i) + i \, e^{i \phi_0} \Gamma_2(s - i) + i\,e^{- i \phi_0} \Gamma_2 (s + i) \right] + i k u \, A(s)~, \\
\end{cases}
\end{equation}
where we have omitted the overline in the Laplace transformed quantities for notational simplicity. By inserting the first two equations of the system (\ref{sm12}) into the third one, simple algebraic manipulations yield the following expression  
\begin{equation}
\label{sm13}
A( s ) = i k v \, \frac{(s - i k w) \, {\rm sin} \, \phi_0 + {\rm cos} \, \phi_0}{Q(s)} \, \gamma_{1,0} + i k v \, \frac{(s - i k w) \, {\rm cos} \, \phi_0 - {\rm sin} \, \phi_0}{Q(s)} \, \gamma_{2,0} + \frac{s^2 - 2 i k w s + 1 - k^2 w^2}{Q(s)} \, a_0~,
\end{equation}
where we have set         
\begin{equation}
\label{sm14}
Q( s ) \equiv s^3 - i k (2 w + u) s^2 + \bigl(1 - k^2 \, w^2 - \, 2k^2 \, u w + k^2 \, v^2 \bigr) s + i k \bigl(k^2 \, w^2 u - u - k^2 \, v^2 w \bigr)~.
\end{equation}
Eq. (\ref{sm13}) can be inverse-Laplace-transformed so as to recover $a(\zeta)={\mathcal L}_+^{-1}\{A\}(\zeta)$. In order to inverse-Laplace-transform $A(s)$ back to the $\zeta$ variable it is useful to recast Eq. (\ref{sm13}) into a different form by using the so-called {\it partial-fraction decomposition}. Correspondingly, we have to distinguish three cases depending on the roots of $Q(s)$. Specifically, $Q(s)$ can have: 1) three distinct roots, 2) two distinct roots (one of them with multiplicity 2), and 3) three coincident roots (one root with multiplicity 3).

\vspace{.2cm}
\noindent {\bf Case A: Q(s) has three distinct roots} $s_1$, $s_2$, $s_3$ -- We can write
\begin{equation}
\label{sm15}
Q ( s ) = (s - s_1)(s - s_2)(s - s_3)~.
\end{equation}
The calculation of $s_1$, $s_2$ and $s_3$ involves well known algebraic manipulations which we do not report here. Our next step consists in rewriting each of the three fractions entering Eq. (\ref{sm13}) in terms of $s_1$, $s_2$ and $s_3$. We start with the 
one multiplied by $\gamma_{1,0}$. Then we must find three constants $\mu_1$, $\mu_2$ and $\mu_3$ which satisfy the relation
\begin{equation}
\label{sm16}
\frac{(s - i k w) \, {\rm sin} \, \phi_0 + {\rm cos} \, \phi_0}{(s - s_1) (s - s_2) ( s - s_3)} = \sum_{n=1}^3 \frac{\mu_n}{s - s_n}~.
\end{equation}
Manifestly, $\mu_1$, $\mu_2$ and $\mu_3$ have to obey the linear system 
\begin{equation}
\label{sm17}
\begin{cases}
\mu_1 + \mu_2 + \mu_3 = 0~, \\[4pt]
\mu_1(s_2 + s_3) + \mu_2 (s_1 + s_3) + \mu_3 (s_1 + s_2) = - \, {\rm sin} \, \phi_0~, \\[4pt]
\mu_1 s_2 s_3 + \mu_2 s_1 s_3 + \mu_3 s_1 s_2 = - \, i k w \, {\rm sin} \, \phi_0 + {\rm cos} \, \phi_0~,
\end{cases}
\end{equation}
whose solution is
\begin{equation}
\label{sm18}
\mu_n = \frac{(s_n - i k w) \, {\rm sin} \, \phi_0 + {\rm cos} \, \phi_0}{(s_n - s_r)(s_n - s_p)}
\end{equation}
with $n, r, p = 1, 2, 3$ and $n \neq r \neq p$. Next, we consider the fraction multiplied by $\gamma_{2,0}$. As before, we have to find three constants $\nu_1$, $\nu_2$ and $\nu_3$ that meet the condition
\begin{equation}
\label{sm16a}
\frac{(s - \, i k w) \, {\rm cos} \, \phi_0 - {\rm sin} \,\phi_0}{(s - s_1) (s - s_2) ( s - s_3)} = \sum_{n=1}^3 \frac{\nu_n}{s - s_n}~.
\end{equation}
Again, $\nu_1$, $\nu_2$ and $\nu_3$ should satisfy the linear system
\begin{equation}
\label{sm17a}
\begin{cases}
\nu_1 + \nu_2 + \nu_3 = 0~, \\[4pt]
\nu_1 (s_2 + s_3) + \nu_2 (s_1 + s_3) + \nu_3 (s_1 + s_2) = - \, {\rm cos} \, \phi_0~, \\[4pt]
\nu_1 s_2 s_3 + \nu_2 s_1 s_3 + \nu_3 s_1 s_2 = - \, i k w \, {\rm cos} \, \phi_0 - {\rm sin} \phi_0~,
\end{cases}
\end{equation}
whose solution is
\begin{equation}
\label{sm18a}
\nu_n = \frac{(s_n - i k w) \, {\rm cos} \, \phi_0 - {\rm sin} \, \phi_0}{(s_n - s_r)(s_n - s_p)}
\end{equation}
with $n, r, p = 1, 2, 3$ and $n \neq r \neq p$. Finally, we address the fraction multiplied by $a_0$. Proceeding as above, our job is  to find three constants $\lambda_1$, $\lambda_2$ and $\lambda_3$ which obey the relation
\begin{equation}
\label{sm16aa}
\frac{s^2 - 2 i k w s + 1 - k^2 w^2}{(s - s_1) (s - s_2) ( s - s_3)} = \sum_{n=1}^3 \frac{\lambda_n}{s - s_n}~.
\end{equation}
Once again, $\lambda_1$, $\lambda_2$ and $\lambda_3$ should obey the linear system
\begin{equation}
\label{sm17aa}
\begin{cases}
\lambda_1 + \lambda_2 + \lambda_3 = 1~, \\[4pt]
\lambda_1 (s_2 + s_3) + \lambda_2 (s_1 + s_3) + \lambda_3 (s_1 + s_2) = 2 i k w~, \\[4pt]
\lambda_1 s_2 s_3 + \lambda_2 s_1 s_3 + \lambda_3 s_1 s_2 = 1 - k^2 w^2~,
\end{cases}
\end{equation}
whose solution is
\begin{equation}
\label{sm18aa}
\lambda_n = \frac{1 + s_n^2 - 2 i k w \, s_n - k^2 w^2}{(s_n - s_r)(s_n - s_p)}
\end{equation}
with $n, r, p = 1, 2, 3$ and $n \neq r \neq p$.

At this point, we are in position to recast Eq. (\ref{sm13}) into the form
\begin{equation}
\label{sm21}
A( s ) = i k v \, \gamma_{1,0} \sum_{n=1}^3 \frac{\mu_n}{s - s_n} + i k v \, \gamma_{2,0} \sum_{n=1}^3 \frac{\nu_n}{s - s_n} + a_0 \sum_{n=1}^3 \frac{\lambda_n}{s - s_n}~.
\end{equation}
In order to get $\overline{\gamma}_1 (\zeta)$ and $\overline{\gamma}_2 (\zeta)$ we should start by inserting Eq. (\ref{sm21}) back into the first and second equations of the system (\ref{sm12}), and then perform the inverse-Laplace transform. But we can save work by proceeding as follows. We first inverse-Laplace-transform Eq. (\ref{sm21}), thereby obtaining
\begin{equation}
\label{sm22}
\overline{a} (\zeta) = {\mathcal L}_+^{-1} \{A\}(\zeta) = i k v \, \gamma_{1,0} \sum_{n=1}^3 \mu_n e^{s_n\zeta} + i k v \, \gamma_{2,0} \sum_{n=1}^3 \nu_n e^{s_n \zeta} + a_0 \sum_{n=1}^3 \lambda_n e^{s_n\zeta}~.
\end{equation}
Next, we observe that the first and second equations of the system (\ref{sm9}) can be rewritten as
\begin{equation}
\label{sm23}
\begin{cases}
\dfrac{d}{d \zeta} \left[\overline{\gamma}_1(\zeta) e^{- i k w \zeta} \right] = i k v \, {\rm sin} (\phi_0 + \zeta) \, e^{- i k w \zeta} \,\overline{a} (\zeta)~,\\[8pt]
\dfrac{d}{d \zeta} \left[\overline{\gamma}_2(\zeta) e^{- i k w \zeta} \right] = i k v \, {\rm cos} (\phi_0 + \zeta) \, e^{- i k w \zeta} \,\overline{a} (\zeta)~.\\[8pt]
\end{cases}
\end{equation}
Hence, we trivially find
\begin{equation}
\label{sm24}
\begin{cases}
\overline{\gamma}_1(\zeta) = e^{i k w \zeta} \left[\gamma_{1,0} + i k v \displaystyle\int_0^\zeta{\,{\rm sin}(\phi_0+\zeta') \, e^{- i k w \zeta'} \,\overline{a}(\zeta') d\zeta'} \right]~,\\[12pt]
\overline{\gamma}_2(\zeta) = e^{i k w \zeta} \left[\gamma_{2,0} + i k v \displaystyle\int_0^\zeta{\,{\rm cos}(\phi_0+\zeta') \, e^{- i k w \zeta'} \,\overline{a}(\zeta') d\zeta'} \right]~.\\[12pt]
\end{cases}
\end{equation}
Before proceeding further, we stress for our subsequent needs that the system (\ref{sm24}) is not explicitly dependent on $Q (s)$. So, we plug Eq. (\ref{sm22}) into the the system (\ref{sm24}). By exploiting the linearity of the integral operator and recalling Eqs. (\ref{sm7}), (\ref{sm8q}) and (\ref{sm8w}) we finally obtain the solution of the system (\ref{sm6}). Whence
\begin{equation}
\label{sm25}
\begin{cases}
{\gamma}_1( y ) = e^{i w (y - y_0)} \left[\gamma_{1,0} \left(1- k^2 v^2 \displaystyle{\sum_{n=1}^3} \mu_n I_n (y - y_0) \right) - \gamma_{2,0} k^2 v^2 \displaystyle{\sum_{n=1}^3} \nu_n I_n (y - y_0) + i a_0 k v \displaystyle{\sum_{n=1}^3}\lambda_n 
I_n ( y - y_0) \right]~, \\[12pt]
{\gamma}_2 ( y ) = e^{i w (y - y_0)} \left[- \, \gamma_{1,0} k^2 v^2 \displaystyle{\sum_{n=1}^3} \mu_n J_n (y - y_0) + \gamma_{2,0} \left(1 - k^2 v^2 \displaystyle{\sum_{n=1}^3} \nu_n J_n (y - y_0) \right) + i a_0 k v \displaystyle{\sum_{n=1}^3} \lambda_n J_n (y - y_0) \right]~, \\[12pt]
{a} ( y ) = i k v \, \gamma_{1,0} \displaystyle{\sum_{n=1}^3} \mu_n e^{s_n k^{-1} (y - y_0)} + i k v \, \gamma_{2,0} \displaystyle{\sum_{n=1}^3} \nu_n e^{s_n k^{-1} (y - y_0)} + a_0 \displaystyle{\sum_{n=1}^3} \lambda_n e^{s_n k^{-1} (y - y_0)}~,
\end{cases}
\end{equation}
where we have set
\begin{eqnarray}
&\displaystyle I_n (y - y_0) \equiv   \label{sm26} \\
&\displaystyle \displaystyle\int_0^{k^{-1}(y-y_0)}{\,{\rm sin}(\phi_0+\zeta') \, e^{(s_n- i k w) \zeta'}d\zeta'} = \frac{e^{(s_n k^{-1} - i w)(y - y_0)} \bigl[s( y ) (s_n - i k w) - c( y ) \bigr] - s(y_0) 
(s_n - i k w) + c( y_0 )}{(s_n - i k w)^2 + 1}~,   \nonumber
\end{eqnarray}
\begin{eqnarray}
&\displaystyle J_n (y - y_0) \equiv   \label{sm27} \\
&\displaystyle \displaystyle\int_0^{k^{-1}(y-y_0)}{\,{\rm cos}(\phi_0+\zeta') \, e^{(s_n- i k w) \zeta'}d\zeta'} = \frac{e^{(s_n k^{-1} - i w) (y - y_0)} \bigl[c( y ) (s_n -i k w) + s( y ) \bigr] - c(y_0)(s_n - i k w ) - s(y_0)}{(s_n - i k w )^2 + 1}~. \nonumber
\end{eqnarray}
The great advantage of Eq. (\ref{sm25}) is that it can easily be rewritten in the form $\psi ( y ) = {\cal U}_{\rm var} \bigl({\cal E}; y, y_0; \phi (y) \bigr) \, \psi(y_0)$, which implies that ${\cal U}_{\rm var} \bigl({\cal E}; y, y_0; \phi (y) \bigr)$ takes the explicit form 
\begin{eqnarray}
&\displaystyle {\cal U}_{\rm var} \bigl({\cal E}; y, y_0; \phi (y) \bigr) = \label{smU} \\         
&\displaystyle \left(\begin{array}{ccc}  
e^{i w (y - y_0)} \left(1 - k^2 v^2 \displaystyle{\sum_{n=1}^3} \, \mu_n I_n (y - y_0) \right) & - k^2 v^2 \, e^{i w (y - y_0)}\displaystyle{\sum_{n=1}^3} \, \nu_n I_n (y - y_0) & i k v \, e^{i w (y-y_0)} \displaystyle{\sum_{n=1}^3} \, \lambda_n I_n (y - y_0) \\[12pt]

- k^2 v^2 e^{i w (y - y_0)} \displaystyle{\sum_{n=1}^3} \, \mu_n J_n (y - y_0) & e^{i w (y - y_0)} \left(1 - k^2 v^2 \displaystyle{\sum_{n=1}^3} \, \nu_n J_n (y - y_0) \right) & i k v \, e^{i w (y - y_0)} \displaystyle{\sum_{n=1}^3} \, \lambda_n J_n (y - y_0) \\[12pt]

i k v \displaystyle{\sum_{n=1}^3} \, \mu_n \, e^{s_n k^{-1} (y - y_0)} & i k v \displaystyle{\sum_{n=1}^3} \, \nu_n \, e^{s_n k^{-1} (y - y_0)} & \displaystyle{\sum_{n=1}^3} \, \lambda_n \, e^{s_n k^{-1} (y - y_0)}  \nonumber
\end{array}
\right)~.
\end{eqnarray}

As far as the case $\phi_1- \phi_0 <0$ is concerned -- in which we must use the anti-causal unilateral Laplace transform ${\mathcal L}_-$ -- there are only two changes in the previous calculation. 

\begin{enumerate}

\item In the system (\ref{sm12}) the replacements $\gamma_{1,0} \to - \, \gamma_{1,0}$, $\gamma_{2,0} \to - \, \gamma_{2,0}$, $a_0 \to - \, a_0$ have to be performed since -- if $f(\zeta)$ is a ${\mathcal L}_{\pm}$-transformable generic function and $F(s)$ is its Laplace transform -- we see that ${\mathcal L}_+ \{ d/d\zeta [f(\zeta)]  \} ( s ) =sF(s)-f(0)$ while ${\mathcal L}_- \{ d/d\zeta [f(\zeta)]  \} ( s ) =s F(s) + f(0)$. The other Laplace transforms in the system (\ref{sm12}) are the same for both ${\mathcal L}_+$ and ${\mathcal L}_-$. As a result, we obtain $A(s) \to - A(s)$ in Eq. (\ref{sm21}).

\item It is easy to see that while ${\mathcal L}_+ \{e^{c \zeta} \} ( s ) = 1/(s - c)$, we have instead ${\mathcal L}_- \{e^{c \zeta}\} ( s ) = - \, 1/(s - c)$ with $c$ a generic constant. 
\end{enumerate}

\noindent Owing to point 2 as applied to Eq. (\ref{sm21}) -- and recalling point 1 -- we see that $\overline{a} (\zeta)$ remains unchanged, and so our previous statement that the solution is the same for both $\phi_1 - \phi_0 >0$ and $\phi_1 - \phi_0 <0$ follows at once. 
          
\

For completeness, we consider below also the two cases in which $Q(s)$ has two or three coincident roots. Basically, the logic of the argument remains the same, and so we briefly stress the points that are different.

\vspace{.2cm}
\noindent {\bf Case B: Q(s) has two distinct roots with one of them with multiplicity 2} -- In this case we can write
\begin{equation}
\label{sm28}
Q( s ) = (s - s_1)(s - s_D)^2~,
\end{equation}
where $s_1$ is the root of $Q(s)$ with multiplicity 1 and $s_D$ is the root with the multiplicity 2. The calculation of $s_1$ and $s_D$ are straightforward. Once they have been determined, we evaluate separately the three fractions in Eq. (\ref{sm13}). 
Starting from the fraction multiplied by $\gamma_{1,0}$ we should find three constants $\mu_1$, $\mu_2$ and $\mu_3$ satisfying the condition
\begin{equation}
\label{sm29}
\frac{(s - i k w) \, {\rm sin} \, \phi_0 + {\rm cos} \, \phi_0}{(s - s_1)(s-s_D)^2} = \frac{\mu_1}{s - s_1} + \frac{\mu_2}{s - s_D} +\frac{\mu_3}{(s - s_D)^2}~.
\end{equation}
The calculation of $\mu_1$, $\mu_2$ and $\mu_3$ proceeds just as in case A. Just the same technique ought to be applied to the fractions multiplied by $\gamma_{2,0}$ and $a_0$ in Eq (\ref{sm13}). Knowing the coefficients entering the partial fraction decomposition we can write an expression similar to  Eq. (\ref{sm21}). Here, the only difference is that we have a term $\propto 1/(s-s_D)^2$ whose inverse Laplace transform is ${\mathcal L}_+^{-1}\{ 1/(s-s_D)^2\} (\zeta) = \zeta \, e^{s_D \zeta}$. With this information we can compute $\overline{a} (\zeta)$ by inverse-Laplace-transforming $A (s)$. Then, the calculation proceeds as in case A: we insert $\overline{a} (\zeta)$ into the system (\ref{sm24}) -- which we know to be  not directly dependent of $Q(s)$ -- and by carrying out the integration we get $\gamma_1 (y)$ and $\gamma_2 (y)$ along with $a (y)$, thanks to the change of variable (\ref{sm7}). Concerning the situation $\phi_1- \phi_0 < 0$, it can be shown that ${\mathcal L}_-^{-1}\{-1/(s - s_D)^2\}( \zeta ) = \zeta \, e^{s_D \zeta}$, and so everything we stated in the case A above holds true even here, thereby implying that the solution is the same for both $\phi_1- \phi_0 > 0$ and $\phi_1- \phi_0 < 0$.

\vspace{.2cm}
\noindent {\bf Case C: Q(s) has three coincident roots (one root with multiplicity 3)} -- Presently we can write
\begin{equation}
\label{sm28}
Q(s) = (s - s_T)^3~,
\end{equation}
where $s_T$ is the only one root of $Q(s)$ (with multiplicity 3). The calculation of $s_T$ involves simple algebra and we do 
not report it here. We next evaluate separately the three fractions in Eq (\ref{sm13}) multiplied by $\gamma_{1,0}, \gamma_{2,0}, a_0$. For the first one we have to find the three constants $\mu_1$, $\mu_2$ and $\mu_3$ satisfying
\begin{equation}
\label{sm29}
\frac{(s-ikw) \, {\rm sin} \, \phi_0 + {\rm cos} \, \phi_0}{(s - s_T)^3} = \frac{\mu_1}{s - s_T} + \frac{\mu_2}{(s - s_T)^2} + \frac{\mu_3}{(s - s_T)^3}~.
\end{equation}
This step resembles very closely that in the previous cases, and the same is true for the other two fractions. In this way we are led to an expression for $A (s)$ quite similar to Eq. (\ref{sm21}). The only new thing here is that we have a term $\propto 1/(s-s_T)^3$ whose inverse Laplace transform is ${\mathcal L}_+^{-1}\{ 1/(s - s_T)^3\}(\zeta) = (1/2) \zeta^2 \, e^{s_T\zeta}$. With this information we can evaluate $\overline{a} (\zeta)$ by inverse-Laplace-transforming $A(s)$. Then the rest of the calculation proceeds as in the cases A and B. As far as the situation $\phi_1-\phi_0<0$ is concerned, it is easy to see that ${\mathcal L}_-^{- 1}\{-1/( s - s_T)^3\} (\zeta) = (1/2) \zeta^2 \, e^{s_T\zeta}$ so that everything we stated above holds true both for $\phi_1-\phi_0>0$ and $\phi_1-\phi_0<0$.

\section{Discussion}

Manifestly, at this stage our proposed DLSME magnetic model is fully defined, apart from the probabi\-li\-ty density of the domain size which depends on the specific nature of 
${\cal T}$. The photon survival probability $P^{\rm ALP}_{\gamma \to \gamma} \bigl({\cal E};y_{\rm ex}, \rho_{\rm ex}; y_{\rm in}, \rho_{\rm in}; \{\phi_n\}_{1 \le n \le N} \bigr)$ along a single arbitrary realization of the beam propagation inside the astronomical system ${\cal T}$ is given by Eq. (\ref{12042018a}), where Eq. (\ref{0105201b}) is expressed in terms of the following transfer matrices of the reduced Schr\"odinger-like equation (\ref{redeqprop}): ${\cal U}_{\rm const} \bigl({\cal E}; y_{\rm ex}, y_{1, N - 1}; \phi_{0, N} \bigr)$, ${\cal U}_{\rm const} \bigl({\cal E}; y_{0, n}, y_{1, n - 1}; \phi_{0, n} \bigr)$ given by Eq. (\ref{mravvq2abcapp}) and ${\cal U}_{\rm var} \bigl({\cal E}; y_{1, n}, y_{0, n}; \phi_{1,n}, \phi_{0,n} \bigr)$ provided by Eq. (\ref{smU}).

\begin{figure}[h]     
\begin{center}
\includegraphics[width=0.50\textwidth]{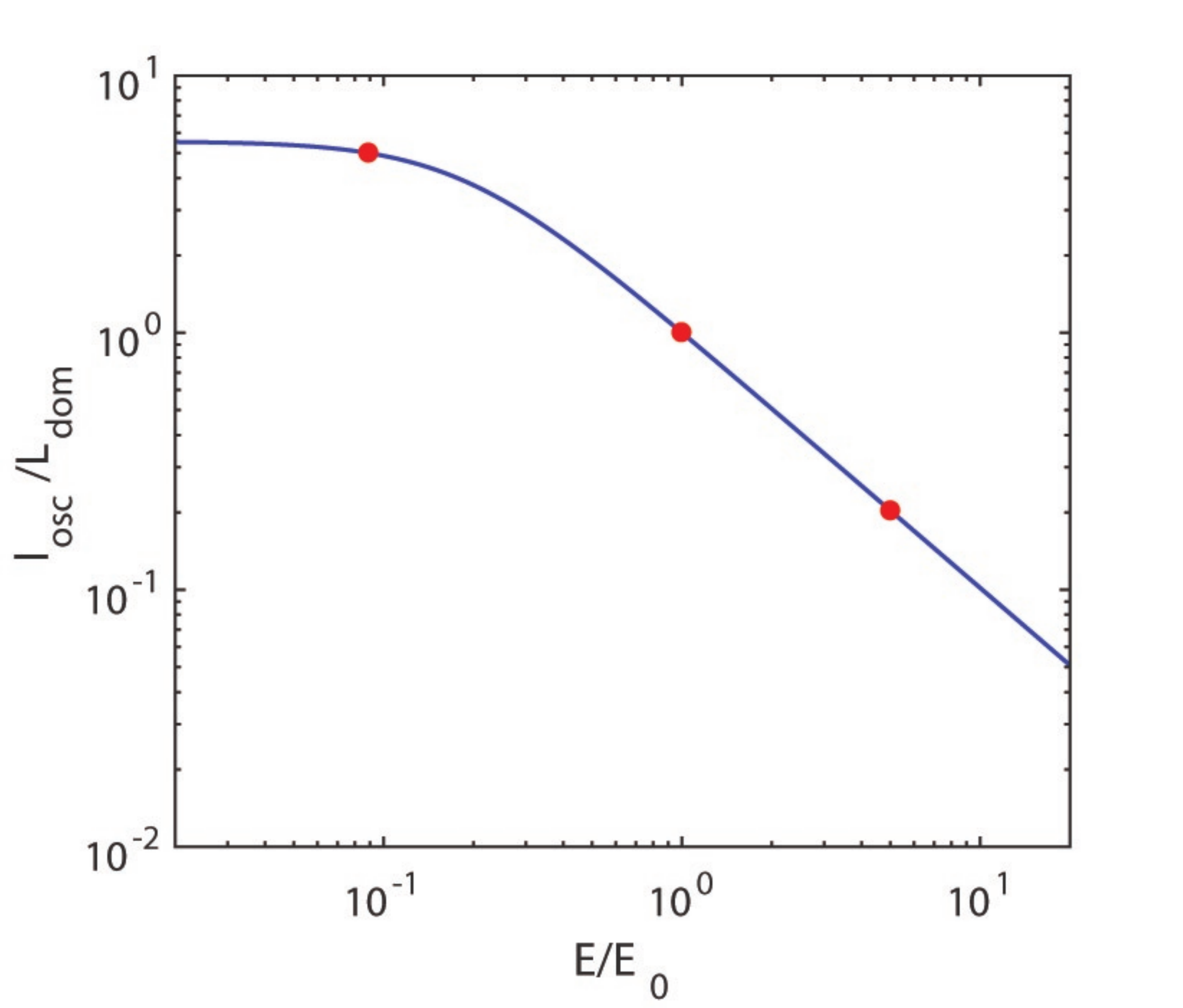}
\end{center}
\caption{\label{Losc} Behavior of $l_{\rm osc} ({\cal E})/L_{\rm dom}$ versus ${\cal E}/{\cal E}_0$ where ${\cal E}_0$  is the energy where  $l_{\rm osc} ({\cal E}_0) \simeq L_{\rm dom}$. The left blob corresponds to $l_{\rm osc} \simeq 5 \, L_{\rm dom}$, the central blob corresponds to $l_{\rm osc} = L_{\rm dom}$ and the right blob corresponds to $l_{\rm osc} \simeq 0.2 \, L_{\rm dom}$.}
\end{figure}

We proceed to investigate how much the shape of the magnetic domains -- which is quantified by the smoothing parameter $\sigma$ (defined in Section V) -- affects $P^{\rm ALP}_{\gamma \to \gamma} \bigl({\cal E};y_{\rm ex}, \rho_{\rm ex}; y_{\rm in}, \rho_{\rm in}; \{\phi_n\}_{1 \le n \le N} \bigr)$ as ${\cal E}$ varies, assuming for simplicity that all domains have equal length $L_{\rm dom}$. Actually, Figure~\ref{LoscComp} shows that the behavior of $l_{\rm osc} ({\cal E})$ is symmetric at low energies ($\propto {\cal E}$) and at high energies ($\propto 1/{\cal E}$). Just the same trend concerns $P_{\gamma \to a} ({\cal E}, L_{\rm dom})$, as it is evident from Figure~\ref{ConvPr}. Therefore, we will focus our attention on what happens starting with a constant value of $l_{\rm osc} ({\cal E})$ in the strong mixing regime, going next towards higher energies where $l_{\rm osc} ({\cal E})$ eventually decreases. Consequently, we choose the parameters entering the mixing matrix (\ref{sm3}) in such a way that as ${\cal E}$ -- supposed initially to be larger than ${\cal E}_L$ -- further increases and $l_{\rm osc} ({\cal E})$ starts out constant with $l_{\rm osc} > L_{\rm dom}$ and retains such a constant behavior: so at the beginning the QED contribution is negligible, consistently with the strong-mixing regime. Yet, as ${\cal E}$ gets larger and larger the QED contribution becomes dominant and $l_{\rm osc} ({\cal E})$ starts to decrease (this will indeed be the situation addressed elsewhere~\cite{paper2}). It is convenient to denote by ${\cal E}_0$ the energy such that  $l_{\rm osc} ({\cal E}_0) = L_{\rm dom}$. Hence, as long as ${\cal E} < {\cal E}_0$ the sinus in Eq. (\ref{a18}) can be expanded and correspondingly the DLSHE model becomes viable (recall that we recover ${\rm DLSHE \, model}$ from the ${\rm DLSME \, model}$ as $\sigma \to 0$). Moreover, in order to gain in clarity it is also convenient to express $l_{\rm osc} ({\cal E})$ in units of the domain length $L_{\rm dom}$ and likewise to measure ${\cal E}$ in units of ${\cal E}_0$. Such a behavior -- which is independent of $\sigma$ -- is shown in Figure~\ref{Losc}.

\begin{figure}[h]      
\begin{center}
\includegraphics[width=0.50\textwidth]{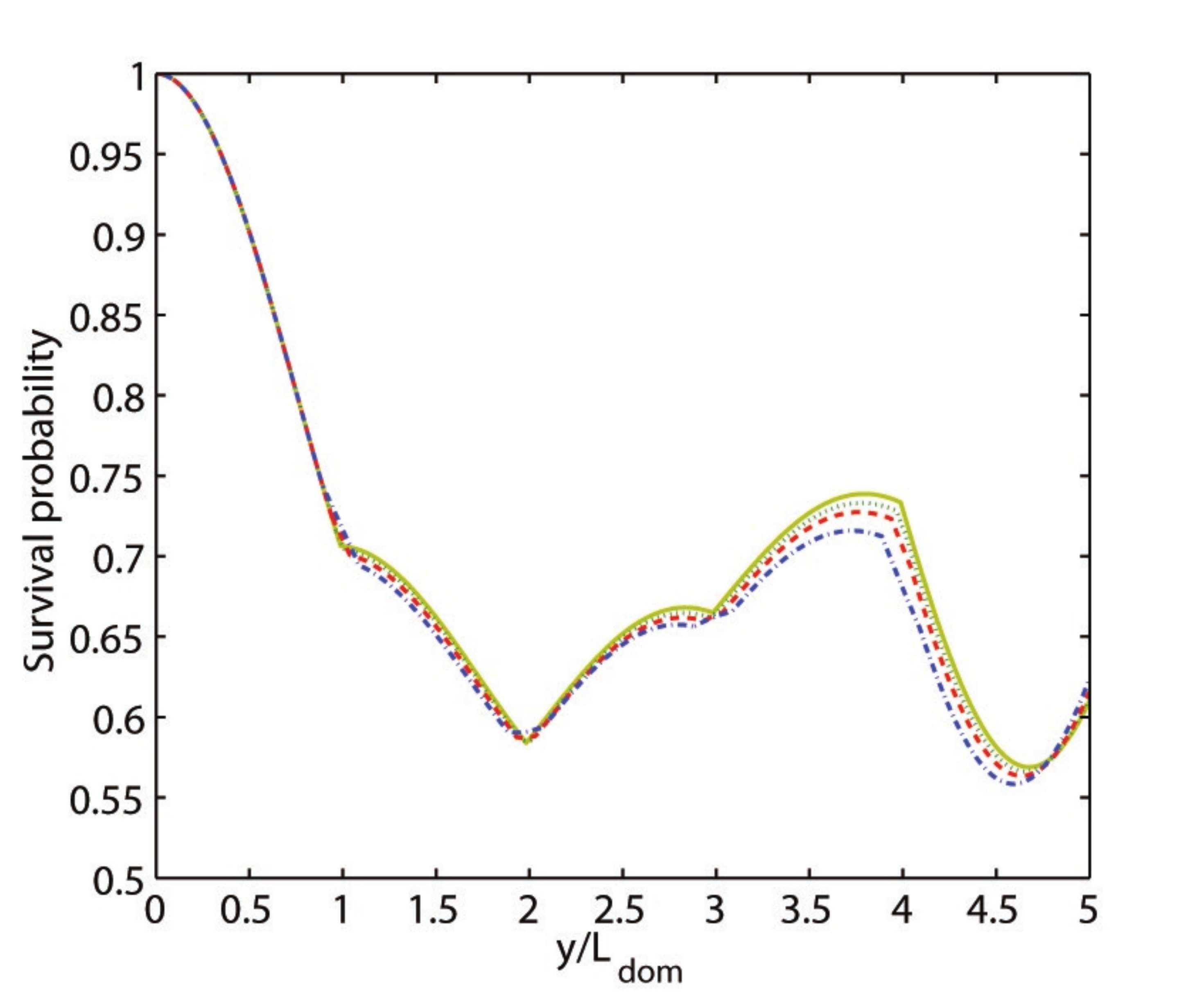}\\
\end{center}
\caption{\label{PvsDist1} Plot of $P^{\rm ALP}_{\gamma \to \gamma} \bigl({\cal E}; y_{\rm ex}, \rho_{\rm ex}; y_{\rm in}, \rho_{\rm in}; \{\phi_n\}_{1 \le n \le N} \bigr)$ versus the distance $y \equiv y_{\rm ex} - y_{\rm in}$ as measured in units of $L_{\rm dom}$ upon variation of the smoothing parameter $\sigma$. The energy is kept fixed and corresponds to the left blob in Figure~\ref{Losc}, namely ${\cal E} \simeq 10^{-1} {\cal E}_0$. We have taken suitable values for the parameters entering the matrix of Eq. (\ref{sm3}) in order to show the effect of the smoothing procedure. This Figure corresponds to the case $l_{\rm osc} ({\cal E}) \simeq 5 \, L_{\rm dom}$.  The solid (yellow) line corresponds to the sharp transition among the magnetic domains ($\sigma =0$), the dotted (green) line to $\sigma =0.05$, the dashed (red) line to $\sigma =0.1$, the dotted-dashed (blue) line to $\sigma =0.2$.}
\end{figure}

\begin{figure}[h]      
\begin{center}
\includegraphics[width=0.50\textwidth]{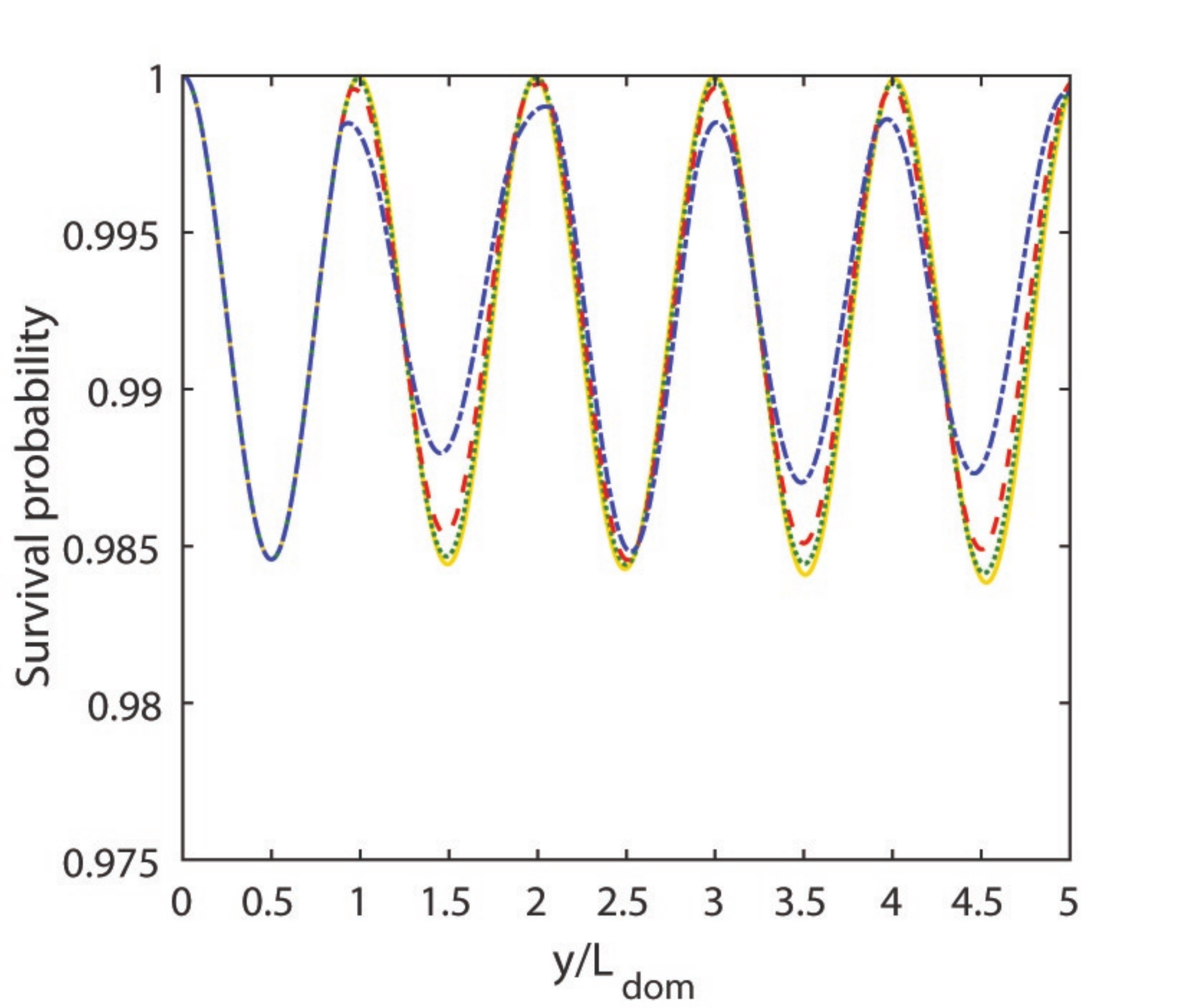}
\end{center}
\caption{\label{PvsDist2} 
Same as Figure~\ref{PvsDist1} but with $l_{\rm osc} ({\cal E}) \simeq L_{\rm dom}$. The energy corresponds to the central blob in Figure~\ref{Losc}, namely ${\cal E} \simeq {\cal E}_0$.}
\end{figure}

\begin{figure}[h]      
\begin{center}
\includegraphics[width=0.50\textwidth]{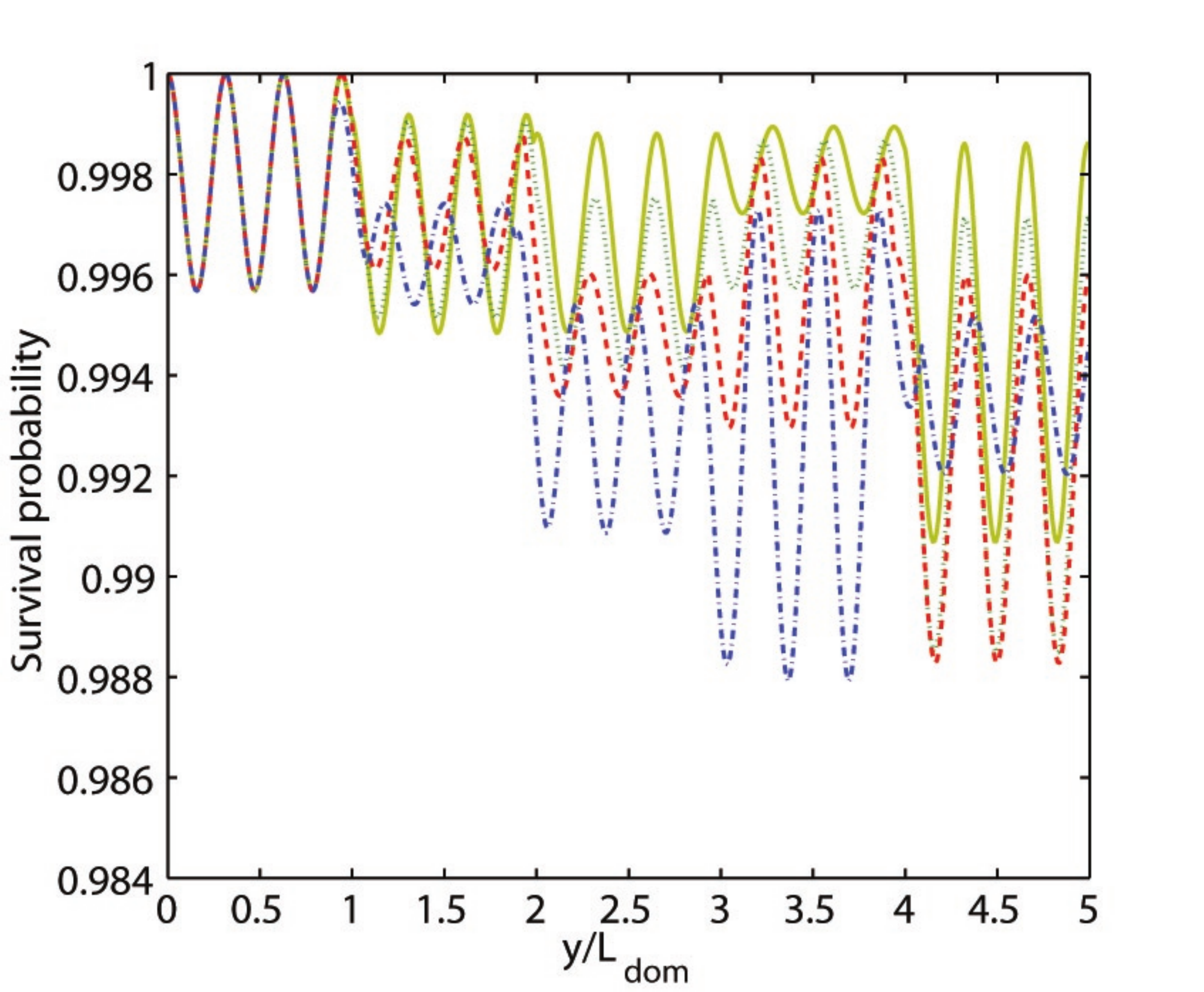}
\end{center}
\caption{\label{PvsDist3} 
Same as Figure~\ref{PvsDist1} but with $l_{\rm osc} ({\cal E}) \simeq 0.2 \, L_{\rm dom}$. The energy corresponds to the right blob in Figure~\ref{Losc}, namely ${\cal E} \simeq 5 \, {\cal E}_0$.}
\end{figure}

Let us next vary the smoothing parameter $\sigma$ so as to find out how $P^{\rm ALP}_{\gamma \to \gamma} \bigl({\cal E};y_{\rm ex}, \rho_{\rm ex}; y_{\rm in}, \rho_{\rm in}; \{\phi_n\}_{1 \le n \le N} \bigr)$ is sensitive to the shape of the magnetic domains at {\it fixed} energy. We plot in Figures~\ref{PvsDist1},~\ref{PvsDist2} and~\ref{PvsDist3} $P^{\rm ALP}_{\gamma \to \gamma} \bigl({\cal E}; y_{\rm ex}, \rho_{\rm ex}; y_{\rm in}, \rho_{\rm in}; \{\phi_n\}_{1 \le n \le N} \bigr)$ versus the distance $y \equiv y_{\rm ex} - y_{\rm in}$ as measured in units of $L_{\rm dom}$ upon variation of the smoothing parameter $\sigma$ for different $l_{\rm osc} ({\cal E})/L_{\rm dom}$ ratios, which are represented by the blobs in  Figure~\ref{Losc}. They all refer to 5 domains, under the assumption that only photons are emitted by the source. The energy is kept fixed in each of these Figures, but is different in different Figures. The reason why we deal with a single realization of the beam propagation is to get rid of the effects of the averaging procedure over the $\{\phi_n \}_{1 \leq n \leq N}$ angles, since we know that what is observed is only a single realization.

\begin{itemize}

\item Figure~\ref{PvsDist1} refers to the case $l_{\rm osc} \bigl(10^{- 1} \, {\cal E}_0 \bigr) \simeq 5 \, L_{\rm dom}$: we note that in this case only a small fraction of an oscillation lies inside a single domain, and so it does not `see' the shape of the magnetic field. As a consequence the amount of smoothing does not practically affect $P^{\rm ALP}_{\gamma \to \gamma} \bigl({\cal E}; y_{\rm ex}, \rho_{\rm ex}; y_{\rm in}, \rho_{\rm in}; \{\phi_n\}_{1 \le n \le N} \bigr)$. 

\item Figure~\ref{PvsDist2} refers to the case $l_{\rm osc} ({\cal E}_0) \simeq L_{\rm dom}$: now the smoothing procedure gives a sizable change of the shape of the oscillations -- which shows up in $P^{\rm ALP}_{\gamma \to \gamma} \bigl({\cal E}; y_{\rm ex}, \rho_{\rm ex}; y_{\rm in}, \rho_{\rm in}; \{\phi_n\}_{1 \le n \le N} \bigr)$ -- but the effect is not dramatic. 

\item Figure~\ref{PvsDist3} shows instead that in the case $l_{\rm osc} (5 \, {\cal E}_0)  \simeq 0.2 \, L_{\rm dom}$ the smoothing effect is drastic. Now the structure of the magnetic field strongly affects the pattern of the oscillations, with in turn reverberates on $P^{\rm ALP}_{\gamma \to \gamma} \bigl({\cal E}; y_{\rm ex}, \rho_{\rm ex}; y_{\rm in}, \rho_{\rm in}; \{\phi_n\}_{1 \le n \le N} \bigr)$. 

\end{itemize}

As expected, the case $\sigma=0.05$ is not much different as compared to the sharp-edges  case ($\sigma=0$). As the smoothing parameter increases $P^{\rm ALP}_{\gamma \to \gamma} \bigl({\cal E}; y_{\rm ex}, \rho_{\rm ex}; y_{\rm in}, \rho_{\rm in}; \{\phi_n\}_{1 \le n \le N} \bigr)$ becomes more and more smoothing-dependent. In Figure~\ref{PvsDist3} it is possible to recognize five zones where the oscillation of a single realization is coherent: they correspond to the five domains crossed by the beam. It is important to realize a rather subtle effect, which is in line with the physical intuition. In all cases considered above -- consisting of 5 domains -- in the first part of the first domain $P^{\rm ALP}_{\gamma \to \gamma} \bigl({\cal E};y_{\rm ex}, \rho_{\rm ex}; y_{\rm in}, \rho_{\rm in}; \{\phi_n\}_{1 \le n \le N} \bigr)$ behaves in the same fashion regardless of the values of the smoothing parameter $\sigma$ and for all $l_{\rm osc} ({\cal E})/L_{\rm dom}$ ratios: this is due to the fact that the propagation of a 
realization in the first domain starts out with the same constant value of the orientation angle. But the more the realization approaches the end of the first domain and the more the result depends on the value of the smoothing parameter.

\begin{figure}[h]
\begin{center}
\includegraphics[width=0.85\textwidth]{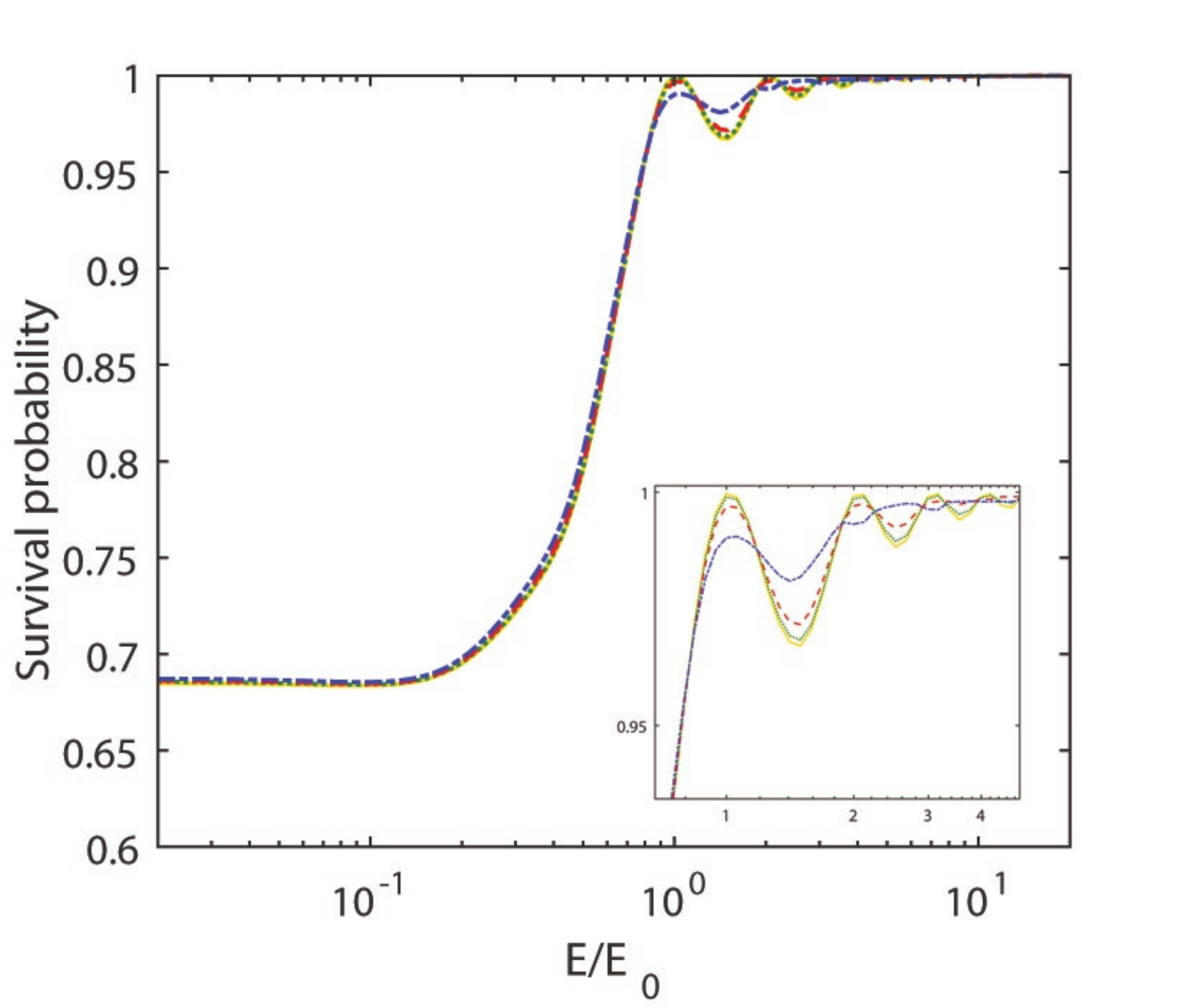}
\end{center}
\caption{\label{AvPvsE}
Behavior of the average unpolarized photon survival probability $P^{\rm ALP, {\rm av}}_{\gamma \to \gamma, {\rm unp}} \bigl({\cal E}; y_{\rm ex}, \rho_x, \rho_z; y_{\rm in}, \rho_{\rm unp} \bigr)$ versus the energy ${\mathcal E}$ as measured in units of ${\mathcal E}_0$. We compute $P^{\rm ALP, {\rm av}}_{\gamma \to \gamma, {\rm unp}} \bigl({\cal E}; y_{\rm ex}, \rho_x, \rho_z; y_{\rm in}, \rho_{\rm unp} \bigr)$ after five domains crossed by the beam upon variation of the smoothing parameter $\sigma$ and for 1000 realizations for every $\sigma$ value. The solid (yellow) line corresponds to the sharp transition among the magnetic domains ($\sigma =0$), the dotted (green) line to $\sigma =0.05$, the dashed (red) line to $\sigma =0.1$, the dotted-dashed (blue) line to $\sigma =0.2$.}
\end{figure}

Finally, it is instructive to consider the energy-dependence of the {\it average} photon survival probability $P^{\rm ALP, {\rm av}}_{\gamma \to \gamma, {\rm unp}} \bigl({\cal E}; y_{\rm ex}, \rho_x, \rho_z; y_{\rm in}, \rho_{\rm unp} \bigr)$ -- as averaged over all unknown angles $\{\phi_n\}_{1 \le n \le N}$ for an unpolarized beam
\begin{eqnarray}
&\displaystyle P^{\rm ALP, {\rm av}}_{\gamma \to \gamma, {\rm unp}} \bigl({\cal E}; y_{\rm ex}, \rho_x, \rho_z; y_{\rm in}, \rho_{\rm unp} \bigr) = \label{12042018b} \\
&\displaystyle \sum_{i = x,z} \left\langle {\rm Tr} \left[\rho_i \, {\cal U}_{\cal T}  
\bigl({\cal E}; y_{\rm ex}, y_{\rm in}; \{ \phi_n \}_{1 \le n \le N} \bigr) \, \rho_{\rm unp} \, {\cal U}_{\cal T}^{\dagger} \bigl({\cal E}; y_{\rm ex}, y_{\rm in}; \{ \phi_n \}_{1 \leq n \leq N} \bigr) \right] \right\rangle_{\{\phi_n\}_{1 \le n \le N}}~, \nonumber
\end{eqnarray}
with $\rho_x$, $\rho_z$, $\rho_{\rm unp}$ defined in Eq. (\ref{sm3a}). Its energy behavior is plotted in Figure~\ref{AvPvsE} for 5 domains, measuring again ${\cal E}$ in units of ${\cal E}_0$. This figure confirms what we got to know from the previous Figures: for low enough energies -- where $l_{\rm osc} ({\cal E}) > \, L_{\rm dom}$ -- there is no difference between the DLSHE and the DLSME models, thereby implying that the smoothing is totally irrelevant. Yet, things begin to become different when the energy starts to be large enough: as $l_{\rm osc} ({\cal E})$ becomes smaller than $L_{\rm dom}$ the average photon survival probability in Eq. (\ref{12042018b}) gets more sensitive to the smoothing parameter $\sigma$. But it is clear from Figure~\ref{AvPvsE} that this probability becomes  again more and more insensitive to the smoothing parameter as the energy further increases, and consequently $l_{\rm osc} ({\cal E})$ gets smaller and smaller. Even this behavior is in agreement with the physical intuition. Indeed, in the intermediate situation the oscillations length is still large enough to show up also in the average, but as the energy further increases the oscillation  length becomes so small that the average procedure washes out its effects altogether.

\section{Applications}

Our investigation has been mainly motivation by the need to cope with the demand posed by the new generation of gamma-ray observatories, which investigate the gamma-ray energy range up to ${\cal E} = 100 \, {\rm TeV}$ and in some case even well beyond. 

While a systematic exploitation of the results obtained here to that field of research will be deferred to a subsequent paper~\cite{paper2}, here we would like to focus our attention on some related aspects. 

In the first place, we would like to stress that we have found a new result: the DLSHE model can yield unphysical results also in the weak mixing regime. Actually, it looks surprising that this fact has gone unnoticed so far, since it is rather obvious from Eq. (\ref{18042018c}) that for low enough energy $l_{\rm osc} ({\cal E})$ can become smaller than $L_{\rm dom}$. 

Let us next come back again to the VHE photon/ALP beam of energy ${\cal E}$ emitted by a blazar, which can be at a very high distance from us. As already pointed out, because of the latter fact it can well happen that it crosses -- apart from the host galaxy and the Milky Way -- some galaxy or galaxy cluster which occasionally lies on the line of sight. Since these are magnetized objects, they can induce additional $\gamma \leftrightarrow a$  oscillations in the beam (beyond those triggered by ${\bf B}_{\rm ext}$). In addition, if the blazar is a flat spectrum radio quasar (FSRQ) the emitted beam will cross the radio lobes. Therefore, it is of importance to know which kind of domain-like model should be used in those cases. A somewhat cursory analysis of the ratio 
$l_{\rm osc} ({\cal E})/L_{\rm dom}$ leads to the results reported in Table~\ref{tabAstrObj}. We have not included clusters of galaxies because they cannot be described by a domain-like magnetic model of the kind considered in the present paper.

\begin{table}[h]
\begin{center}
\begin{tabular}{c|c|c|c|c}
\hline

Astronomical object & \ \  $B/\mu{\rm G}$ \ \ & \ \  $L_{\rm dom}/{\rm kpc}$ \ \ & \ \  Min ${\cal E}/{\rm MeV} {\rm \, such \, that \,} l_{\rm osc} \gtrsim L_{\rm dom}$ & \ \  Max ${\cal E}/{\rm TeV} {\rm \, such \, that \,} l_{\rm osc} \gtrsim L_{\rm dom}$  \ \ \\
\hline
\hline

Radio lobes & 10 & 10 & 1.36 & 690 \\
Spiral galaxies & 7 & 10 & 1.30 & 1350  \\
Starburst galaxies & 50 & 10 & DLSME model always needed & DLSME model always needed  \\
Elliptical galaxies & 5 & 0.15 & $1.88 \cdot 10^{-2}$ & $1.5 \cdot 10^5$   \\
\hline
\end{tabular}
\caption{Various astronomical objects with typical values of average $B$ and $L_{\rm dom}$ that are amenable of a treatment with the domain-like models considered in this paper and which can be crossed by the considered VHE photon/ALP beam. The last two columns define the energy range inside which the DLSHE can be employed ($l_{\rm osc} \gtrsim L_{\rm dom}$). Outside this energy range -- below the energy reported in the fourth column and above the energy in the fifth column -- only the DLSME model can be used in order to get physically sensible results in calculating the photon-ALP propagation.}
\label{tabAstrObj}
\end{center}
\end{table}

Recalling the polarimetric satellite missions IXPE and XIPE as well as the 
conventional/polarimetric e-ASTROGRAM and AMEGO missions, we prefer to leave the initial $\rho_{\rm in}$ and final $\rho_{\rm ex}$ polarizations unspecified. Therefore, the total transfer matrix is still given by Eq. (\ref{0105201b}), but the photon survival probability along a single realization takes the form
\begin{eqnarray}
\label{mr051017bz}
&\displaystyle P^{\rm ALP}_{\gamma \to \gamma} \bigl({\cal E};y_{\rm ex}, \rho_{\rm ex}; y_{\rm in}, \rho_{\rm in}; \{\phi_n\}_{1 \le n \le N} \bigr) = \\
&\displaystyle {\rm Tr} \left[\rho_{\rm ex} \, {\cal U}_{\rm const} \bigl({\cal E}; y_{\rm ex}, y_{\rm in}; \{\phi_n \}_{1 \le n \le N} \bigl) \, \rho_{\rm in} \, {\cal U}^{\dagger}_{\rm const} 
\bigl({\cal E}; y_{\rm ex}, y_{\rm in}; \{\phi_n \}_{1 \le n \le N} \bigr) \right]~. \nonumber
\end{eqnarray}

Which kind of domain-like model for ${\bf B}_{\rm ext}$ should be adopted in this case?  
Taking $g_{ \gamma \gamma a} = {\cal O} (10^{-11} \, \rm GeV^{-1})$,  ${B}_{\rm ext} = {\cal O} (1 \, \rm nG)$ and $L_{\rm dom} = {\cal O} (1 \, {\rm Mpc})$ as benchmark values as an orientation, we find it compelling to use the DLSME model for ${\cal E} \lesssim \, 0.1 \rm GeV$.  

\section{Conclusions}

The aim of the present paper is to present a new, realistic model for the magnetic fields 
which can be described by a domain-like network. So far, a domain-like model with sharp edges has been systematically employed, which possesses the drawback that the magnetic field jumps discontinuously from one domain to the next: needless to say, this sort of model can only be regarded as a highly mathematical idealization. It works only under the unstated assumption that the oscillation length is considerably larger than the domain size. 

Our model avoids such a restriction and holds true even when the oscillation length becomes smaller than the domain size, since we have smoothed out the edges in such a way that the magnetic field varies in a continuous fashion from one domain to the next. As a consequence, it can be applied whenever a magnetic field can be approximated by a domain-like network and its strength varies only slightly across the considered astronomical object ${\cal T}$. The price we had to pay is to solve a system of highly nontrivial differential equations governing the photon/ALP beam propagation inside a single domain, but remarkably we have succeeded to accomplish this task by a clever use of the Laplace transform. We have next computed the photon survival probability along a single realization of the beam propagation process -- which is the really observable quantity -- and we have investigated how it varies as a function of the smoothing parameter. We have found that the smoothing effect becomes important as soon as $l_{\rm osc}$ becomes smaller than $L_{\rm dom}$.

Owing to the newly discovered effect of photon dispersion on the CMB~\cite{raffelt2015}, above a certain energy threshold the DLSHE idealized model yields unphysical result, since the oscillation length becomes comparable or even smaller than the domain size: the reason is that in this case the oscillations probe a whole domain. But -- surprisingly -- we have found that a similar situation occurs also in the weak mixing regime for low enough energy.

Actually, we have been prompted to develop our model by the need to go to energies up to $1000 \, {\rm TeV}$, as required by the new generation of gamma-ray detectors like the 
CTA (Cherenkov Telescope Array)~\cite{cta}, HAWC (High-Altitude Water Cherenkov Observatory)~\cite{hawc}, GAMMA-400 (Gamma Astronomical Multifunctional Modular Apparatus)~\cite{g400}, LHAASO (Large High Altitude Air Shower 
Observatory)~\cite{lhaaso} and TAIGA-HiSCORE (Hundred Square km Cosmic Origin Explorer)~\cite{desy}. As we said, the application of the results derived in this paper will be applied to a photon/ALP beam of energy in the range $10 \, {\rm GeV} - 1000 \, {\rm TeV}$ in a subsequent paper concerning the extragalactic space~\cite{paper2}. Finally, we remark in this respect that the great advantage of dealing with an exact evaluation of the photon survival probability is to drastically reduce the computation time in the applications requiring computer simulations as compared to a numerical solution of the beam propagation equation. 

\section*{Acknowledgments}

We thank Italo Guarneri and Alessandro Mirizzi for discussions, Alessandro De Angelis, Luigina Feretti and Giancarlo Setti for a careful reading of the manuscript and Georg Raffelt for discussions and for communicating to us his results about photon dispersion on the CMB~\cite{raffelt2015} prior to publication. G. G. acknowledges contribution from the grant INAF CTA--SKA, `Probing particle acceleration and $\gamma$-ray propagation with CTA and its precursors', while the work of M. R. is supported by an INFN TAsP grant.


\begin{thebibliography}{20}
\bibitem{rem1} Actually, what is really needed is an electromagnetic field, but in view of our further considerations we focus our attention on a magnetic field.
\end{thebibliography}
\end{document}